%% file: main.tex
\documentclass[10pt,twocolumn,letterpaper]{article}

\usepackage[pagenumbers]{iccv} 


\definecolor{iccvblue}{rgb}{0.21,0.49,0.74}
\usepackage[pagebackref,breaklinks,colorlinks,allcolors=iccvblue]{hyperref}
\usepackage{url}
\usepackage[utf8]{inputenc} 
\usepackage[T1]{fontenc}    
\usepackage{booktabs}       
\usepackage{amsfonts}       
\usepackage{nicefrac}       
\usepackage{microtype}      
\usepackage{multirow}
\usepackage{amsmath}
\usepackage{graphicx}
\usepackage{colortbl}
\usepackage{dsfont}
\usepackage[ruled,vlined]{algorithm2e}
\usepackage[accsupp]{axessibility}  

\def\paperID{12552} 
\def\confName{ICCV}
\def\confYear{2025}

\title{SPA: Efficient User-Preference Alignment against Uncertainty in Medical Image Segmentation}

\author{Jiayuan Zhu \thanks{equal contribution} \\
University of Oxford  \\
{\tt\small jiayuan.zhu@eng.ox.ac.uk}
\and Junde Wu \footnotemark[1] \\
University of Oxford  \\
{\tt\small junde.wu@eng.ox.ac.uk}
\and Cheng Ouyang \\
University of Oxford  \\
{\tt\small cheng.ouyang@eng.ox.ac.uk}
\and Konstantinos Kamnitsas \\
University of Oxford  \\
{\tt\small konstantinos.kamnitsas@eng.ox.ac.uk}
\and J. Alison Noble \\
University of Oxford  \\
{\tt\small alison.noble@eng.ox.ac.uk} \\
}

\begin{document}
\maketitle
\input{sec/0_abstract}    
\input{sec/1_intro}

\input{sec/2_related_work}

\input{sec/3_methodology}

\input{sec/4_experiment}
\input{sec/5_conclusion}
{
    \small
    \bibliographystyle{ieeenat_fullname}
    \bibliography{main}
}
\input{sec/X_suppl}

\end{document}

%% file: sec/0_abstract.tex
\begin{abstract}
Medical image segmentation data 
inherently contain uncertainty. This can stem from both imperfect image quality and variability in labeling preferences on ambiguous pixels, which depend on annotator expertise and the clinical context of the annotations. For instance, a boundary pixel might be labeled as tumor in diagnosis to avoid under-estimation of severity, but as normal tissue in radiotherapy to prevent damage to sensitive structures. As segmentation preferences vary across downstream applications, it is often desirable for an image segmentation model to offer user-adaptable predictions rather than a fixed output. While prior uncertainty-aware and interactive methods offer adaptability, they are inefficient at test time: uncertainty-aware models require users to choose from numerous similar outputs, while interactive models demand significant user input through click or box prompts to refine segmentation. To address these challenges, we propose \textbf{SPA}, a new \textbf{S}egmentation \textbf{P}reference \textbf{A}lignment framework that efficiently adapts to diverse test-time preferences with minimal human interaction. By presenting users with a select few, distinct segmentation candidates that best capture uncertainties, it reduces the user workload to reach the preferred segmentation. To accommodate user preference, we introduce a probabilistic mechanism that leverages user feedback to adapt a model's segmentation preference. The proposed framework is evaluated on several medical image segmentation tasks: color fundus images, lung lesion and kidney CT scans, MRI scans of brain and prostate. SPA shows 1) a significant reduction in user time and effort compared to existing interactive segmentation approaches, 2) strong adaptability based on human feedback, and 3) state-of-the-art image segmentation performance across different imaging modalities and semantic labels. Our code is publicly available here: \url{https://github.com/SuperMedIntel/SPA}.
\end{abstract}

%% file: sec/1_intro.tex
\section{Introduction}
\label{sec:intro}
Deep learning-based medical image segmentation has achieved remarkable progress over the past decade \cite{ronneberger_u-net_2015, tajbakhsh_convolutional_2016, chen_transunet_2021}. However, existing approaches often fail when applied to real-world clinical scenarios. A critical challenge is how to handle the inherent uncertainties in medical images \cite{hesamian_deep_2019}. A single medical image may have multiple different valid segmentation results, depending on the labeling criteria for a specific medical context. For example, in glioma detection from brain CT scans, it is often preferred to include surrounding tissue rather than risk missing part of the tumor \cite{ballou_tumor_1997}. In contrast, in radiation therapy for low-grade glioma, undersegmentation is preferred to protect sensitive brain tissue from excessive radiation damage \cite{prasanna_normal_2012}. Therefore, it is essential to develop adaptive methodologies that align segmentation uncertainties with specific labeling preferences for different clinical needs.

\begin{figure*}[hbt!]
    \centering
\includegraphics[width=0.95\linewidth]{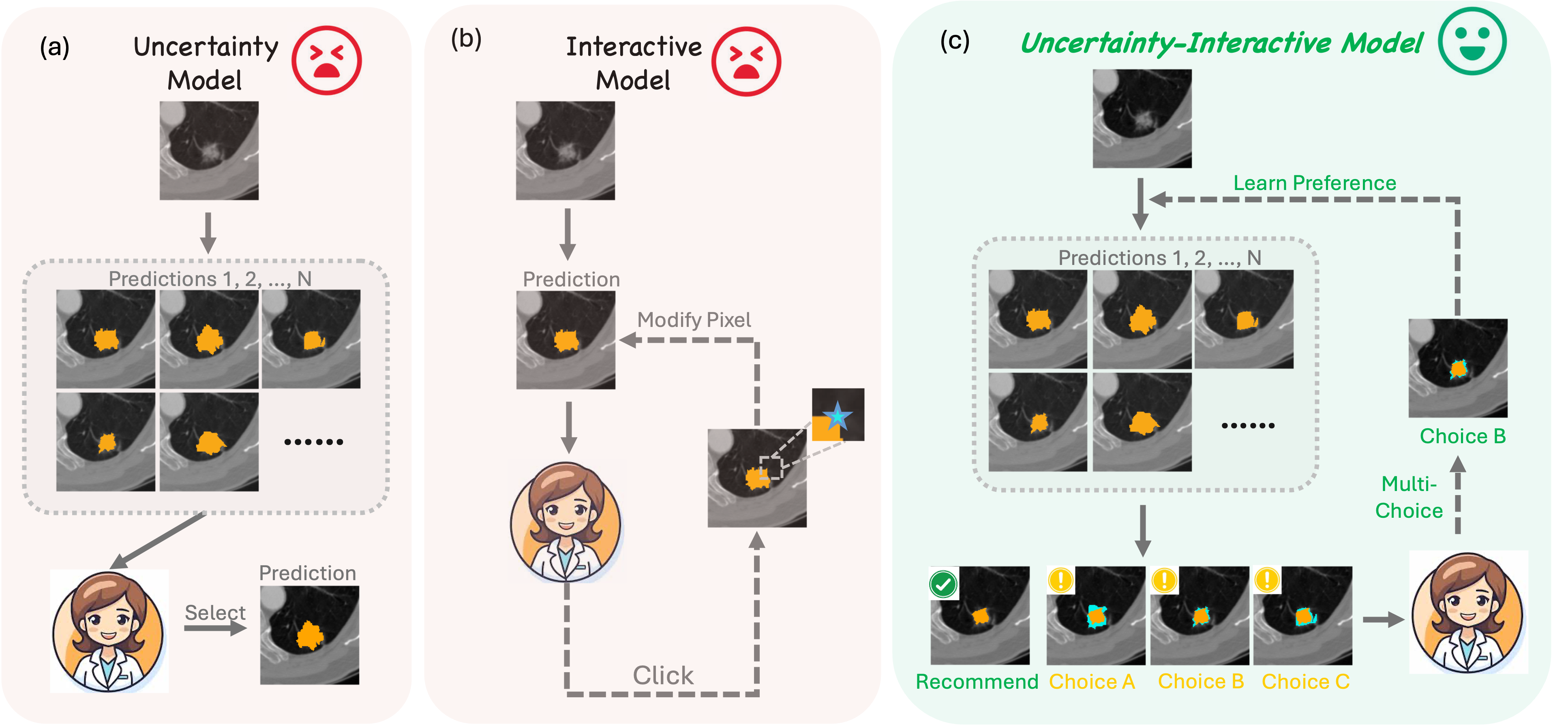}
    \caption{(a) Existing uncertainty-aware models require users to choose from numerous similar-looking candidates, making the process labor-intensive and time-consuming. (b) Interactive segmentation models lack the ability to incorporate image uncertainty and rely heavily on pixel-level user inputs, which require a substantial amount of time and effort. (c) Our uncertainty-aware interactive segmentation model, SPA, efficiently achieves segmentations whose decisions on uncertain pixels are aligned with users preferences. This is achieved by modeling uncertainties and human interactions. At inference time, users are presented with one recommended prediction and a few representative segmentations that capture uncertainty, allowing users to select the one best aligned with their clinical needs. If the user is unsatisfied with the recommended prediction, the model learns from the users' selections, adapts itself, and presents users a new set of representative segmentations. Our approach minimizes user interactions and eliminates the need for painstaking pixel-wise adjustments compared to conventional interactive segmentation models.}
    \label{fig:Workflow}
\vspace{-15pt}
\end{figure*}

Existing uncertainty-aware approaches \cite{kendall_what_2017, rupprecht_learning_2017, baumgartner_phiseg_2019, kohl_probabilistic_2019} represent segmentation uncertainty to users by generating numerous stochastic predictions (Fig. \ref{fig:Workflow} (a)). However, users often have to carefully choose from countless similar-looking candidates, making the process time-consuming. Furthermore, since these models cannot incorporate human feedback for adjustments, there is a risk that none of the predictions be satisfying.

Incorporating human feedback into segmentation, often in the form of visual prompts (such as clicks and bounding boxes) has shown promise \cite{sofiiuk_reviving_2021, chen_focalclick_2022, liu_simpleclick_2023,kirillov_segment_2023}. However, most such interactive segmentation approaches do not incorporate uncertainty. Additionally, existing visual prompts are often cumbersome to attain in the real world, as they require pixel-level interaction input from a user (Fig. \ref{fig:Workflow} (b)).

To address the above challenges, we propose to present users a small number of distinct segmentations to represent uncertainty. It allows users to select their preferred option, simplifying interactive refinement into a straightforward multiple-choice selection. In this paper, we introduce \textbf{SPA}, a new approach for efficient \textbf{S}egmentation \textbf{P}reference \textbf{A}lignment with uncertainty in medical image segmentation (Fig. \ref{fig:Workflow}). SPA presents image uncertainties by generating multiple segmentations. Instead of providing users with numerous similar-looking predictions as in conventional uncertainty-aware segmentations, our model offers one recommended prediction and four representative segmentation candidates per iteration. Once the user selects an option, SPA rapidly adapts through iterative refinement, aligning efficiently with the user's segmentation preference in only a few iterations. Our experiment demonstrates that a user can segment 35\% more images with 39\% fewer iterations compared to previous interactive models. This highlights SPA’s potential for real-world clinical applications. In summary, our contributions are:

\begin{itemize}
\item We introduce SPA, a new segmentation framework that can
align to users' decision preferences toward ambiguous/uncertain pixels at inference time. Such preferences often vary with clinical contexts, which existing segmentation models cannot adapt to efficiently.

\item We propose to model uncertainties in pixel-wise predictions under diverse preferences as a parameterized latent distribution. This modeling enables rapid adaptation to user preferences with minimal iterations at inference time. 

\item We develop a multi-choice interaction mechanism to receive user preferences at inference time, providing a more friendly and effortless user experience than existing works.

\item We compare SPA with deterministic, uncertainty-aware, and interactive SOTA methods. SPA achieves superior results measured by Dice Score for three multi-clinician annotated datasets (REFUGE2, LIDC-IDRI and QUBIQ). SPA also consistently outperforms interactive methods in terms of requiring fewer rounds of user interactions.
\end{itemize}

%% file: sec/2_related_work.tex
\begin{figure*}[hbt!]
    \centering
    \includegraphics[width=0.95\linewidth]{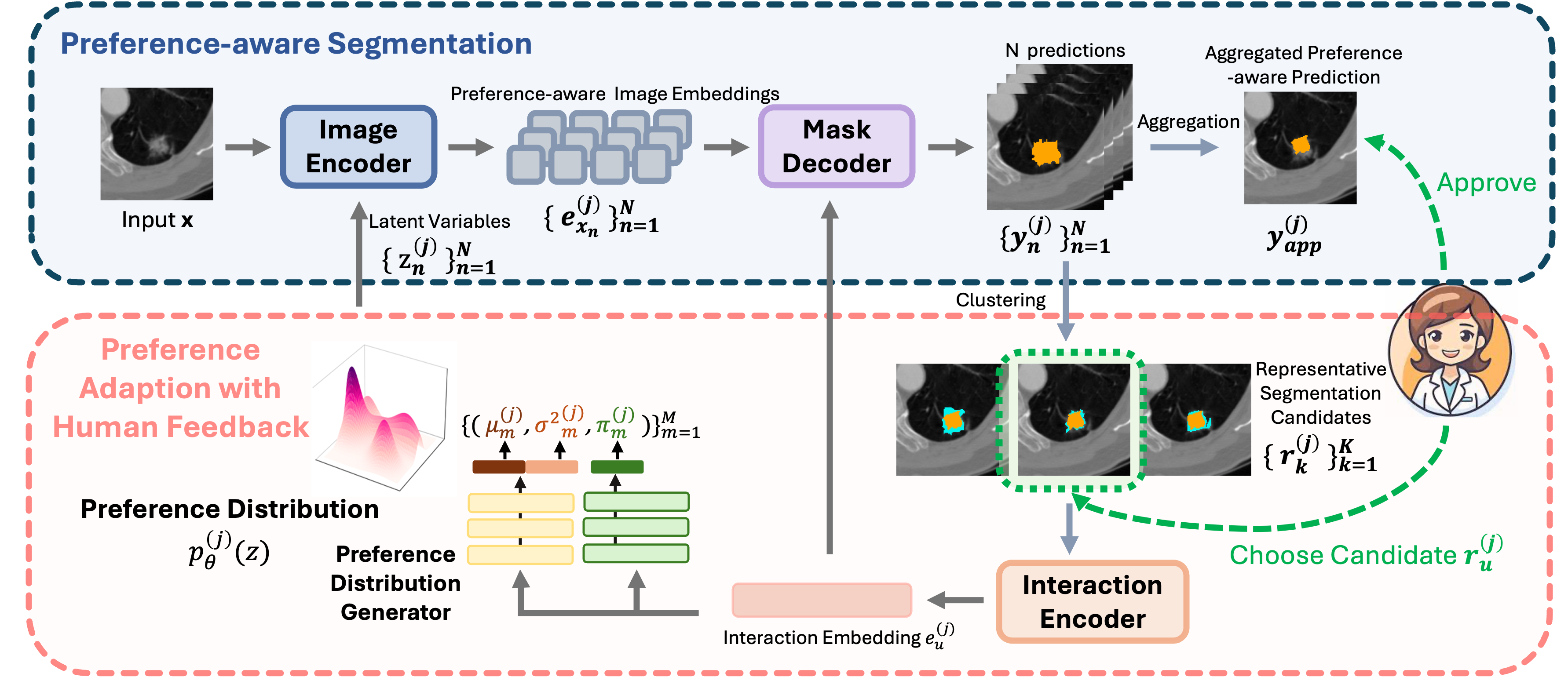} 
    \caption{\textbf{Overall framework of SPA}. The inference process comprises of two steps: \textit{Preference-aware Segmentation} and \textit{Preference Adaptation with Human Feedback}. At iteration $j$, SPA takes the input image $\mathbf{x}$, an interaction embedding $e_u^{(j)}$, and latent variables $\{\mathbf{z_n}^{(j)}\}_{n=1}^{N}$ drawn from the preference distribution $p^{(j)}_{\theta}(z)$ to generate $N$ segmentation predictions. These predictions are then combined into an aggregated preference-aware prediction $\mathbf{{y}_{app}}^{(j)}$. If the user is not satisfied with $\mathbf{{y}_{app}}^{(j)}$, SPA generates $K$ representative segmentation candidates $\{\mathbf{r_k}^{(j)}\}_{k=1}^{K}$ in the adaptation step. The user selects the preferred segmentation candidate $\mathbf{r_u}^{(j)}$. The preference distribution is then updated to $p^{(j+1)}_{\theta}(z)$ based on this choice. This process iterates until the segmentation meets clinical satisfaction.}
    \label{fig:Network}
\vspace{-10pt}
\end{figure*}

\section{Related Work}
\label{sec:related_work}

\noindent
\textbf{Uncertainty-aware Medical Segmentation} Uncertainty in medical images cannot be reduced by adding more data or using more complex models \cite{kiureghian_aleatory_2009}. Techniques such as model ensembling, label sampling \cite{jensen_improving_2019}, and multi-head strategies \cite{guan_who_2018} attempt to address this uncertainty by generating a range of potential predictions that reflect different user preferences \cite{zou_review_2023, huang_review_2024}. Probabilistic segmentation methods, including ProbUNet \cite{kohl_probabilistic_2018}, PhiSeg \cite{baumgartner_phiseg_2019}, CM-Net \cite{zhang_disentangling_2020} and MRNet \cite{ji_learning_2021} explicitly model the posterior distribution of parameters or predictions to capture uncertainty. However, these approaches often generate multiple predictions, requiring users to review each individually. Moreover, the predictions may not perfectly match the specific clinical context, and some techniques rely on prior knowledge of the expertise of a clinician, complicating adoption into clinical practice. In contrast, a more efficient approach is to incorporate human interactions, allowing users to refine segmentations through direct interaction to better align with the clinical context. 

\noindent
\textbf{Interactive Medical Segmentation}
Interactive segmentation is an iterative process where automated segmentation results are refined through user input until they reach a desired output. 
Previous methods \cite{wang_interactive_2018, sakinis_interactive_2019, kirillov_segment_2023} have modified predictions based on user interaction at the pixel level, with some achieving success in the medical domain \cite{deng_sam-u_2023, ma_segment_2024, wu_medical_2023, zhu_medical_2024}. However, these interactive models often do not address the inherent uncertainty in medical images. Additionally, they may require numerous iterations to align with the user preference for a specific clinical context, given the existence of multiple valid segmentations. An interactive model capable of incorporating image uncertainties and learning from user interactions would reduce the number of iterations required to achieve a desired result.

%% file: sec/3_methodology.tex
\section{Methodology}
\subsection{Model user preferences as latent distributions}
We assume that each user's decisions on uncertain/ambiguous pixels, based on their preference and stemming from different medical contexts, can be modeled as \textit{i.i.d.} samples drawn from a parameterized distribution. To represent this user preferences for different clinical contexts, we model this \textit{preference distribution} $p_{\theta}(z)$ as a mixture of $M$ Gaussian distributions:
\begin{equation}
  p_{\theta}(z) = \sum^M_{m = 1} \pi_m \mathcal{N}(z \mid \mu_m, \sigma_m^2)
\vspace{-3pt}
\end{equation} 
where $\mathcal{N}(z \mid \mu_m, \sigma_{m}^2)$ denotes the Gaussian component for user $m$, with mean $\mu_m$ and variance $\sigma_{m}^2$. The mixture weights $\pi_m$ satisfy $\pi_m > 0$ and $\sum_{m=1}^{M}\pi_{m} = 1$. Each component in this mixture model reflects a specific user segmentation preference, enabling the framework to generate segmentation predictions that capture discrepancies across users. In our segmentation framework, $z\sim p_{\theta}(z)$ is implemented as a conditioning signal injected into a neural network, predicating the subtle preference-dependent variability in segmentation outputs.

Modeling user preferences as parameterized distributions is highly adaptable, as the distribution $p_{\theta}(z)$ can be iteratively updated via maximum likelihood based on new observations of user interactions at inference time \footnote{We provide a theoretical proof in the appendix to demonstrate that, given sufficient interactions from a specific user $u$, the preference distribution $p_{\theta}(z)$ will converge to the user's personalized distribution $\mathcal{N}(\mu_u, \sigma_{u}^2)$.}.

We integrate the preference distribution $p_{\theta}(z)$ into the segmentation workflow, as follows. In each iteration, we sample latent variables $\{\mathbf{z_n}\}_{n=1}^N$ from $p_{\theta}(z)$ to capture variations under user preferences \footnote{Preliminary experiments with MRNet \cite{ji_learning_2021} given in the appendix suggest that while segmentation performance varies between users, each user demonstrates consistent results. These observations indicate that modeling interaction behavior can effectively learn user preferences, thereby improving modeling efficiency.}. During training, the model learns $p_{\theta}(z)$ through user feedback simulated from annotated segmentations from multiple users. At inference, the model can efficiently adapt to user preferences by estimating the maximum likelihood based on the feedback of a user.

\subsection{Overall Workflow} \label{ref:train_update}
We propose SPA, which learns to model human preferences by preference distribution $p_{\theta}(z)$ during training, enabling it to efficiently model new user preferences at inference through interaction. Our framework also simplifies and improves the robustness of interaction by replacing pixel-level clicking with multi-choice selection.

Formally, our goal is to learn a general function $f(\cdot, \cdot)$ that can adapt to different user preferences $u$'s through an interaction process: $\mathbf{y} = f(\mathbf{x}, \mathbf{r_u})$, where $\mathbf{x} \in \mathbb{R}^{H \times W \times C}$ is the input image, and $\mathbf{r_u}$ reflects human interaction. The training details are provided in Section \ref{sec:train_details} and Algorithm \ref{alg:preference_gmm}.

Specifically, our SPA framework consists of two main modules (Fig. \ref{fig:Network}): \textit{Preference-aware Segmentation} (Section \ref{section: Space Sampling}) and \textit{Preference Adaption with Human Feedback} (Section \ref{section: Space Updating}). Our \textit{Preference-aware Segmentation} generates multiple valid segmentations to represent image uncertainty (oftentimes \textit{aleatoric}) and the \textit{Preference Adaption with Human Feedback} aligns these segmentations iteratively with specific user preferences. 

In inference, as shown in the upper part of Fig. \ref{fig:Network}, given a raw image $\mathbf{x}$, \textit{Preference-aware Segmentation} generates $N$ segmentation predictions $\{\mathbf{{y}_n}^{(j)}\}_{n=1}^{N}$, conditioned by the interaction embedding $e^{(j)}_u$ representing the user selection, and latent variables $\{\mathbf{z_n}^{(j)}\}_{n=1}^{N}$ sampled from the preference distribution $p_{\theta}^{(j)}(z)$ at iteration $j$. We use the user's first point prompt to produce $e^{0}_u$ and $p_{\theta}^{0}(z)$ for the initialization. Predictions $\{\mathbf{{y}_n}^{(j)}\}_{n=1}^{N}$ reflect uncertainty from individual subtle preferences, which is controlled by $\mathbf{z_n}^{(j)}$. These predictions are combined into one aggregated preference-aware prediction $\mathbf{{y}_{app}}^{(j)}$ and $K$ \textit{Representative Segmentation Candidates} $\{\mathbf{r_k}^{(j)}\}_{k=1}^{K}$. If the user approves $\mathbf{{y}_{app}}^{(j)}$, it will be used as the final segmentation and the iteration ends. Otherwise, the framework uses the \textit{Preference Adaption with Human Feedback} step for further refinement, as shown in the lower part of Fig. \ref{fig:Network}: The user then selects the segmentation $\mathbf{r_u}^{(j)}$ that best aligns with their preference. The preference distribution $p_{\theta}^{(j)}(z)$ is further updated to $p_{\theta}^{(j+1)}(z)$ based on this feedback. This iterative process continues until the user is satisfied with the segmentation.

\subsection{Preference-aware Segmentation} \label{section: Space Sampling}
  As shown in the upper part of Fig. \ref{fig:Network}, our \textit{Preference-aware Segmentation} step involves an image encoder and a mask decoder. Given an input image $\mathbf{x} \in \mathbb{R}^{H \times W \times C}$, we first obtain a general image embedding $\mathbf{e_x}^{(j)} \in \mathbb{R}^{L \times {\frac{H}{P}} \times {\frac{W}{P}}}$ for iteration $j$, where $L$ represents the output channels and $P$ is the Vision Transformer (ViT) patch size. This embedding is generated using a pre-trained ViT \cite{he_masked_2021}. Additionally, we sample $N$ preference conditions $\{\mathbf{z_n}^{(j)}\}_{n=1}^{N}$ from the preference distribution $p_{\theta}^{(j)}(z)$. Each $\mathbf{z_n}^{(j)} \in \mathbb{R}^{L \times {\frac{H}{P}} \times {\frac{W}{P}}}$ is concatenated with the general image embedding $\mathbf{e_x}^{(j)}$, and then processed through three convolutional layers and ReLU activations to produce a set of preference-aware image embeddings $\{\mathbf{e_{x_n}}^{(j)} \}_{n=1}^{N}$, where each $\mathbf{e_{x_n}}^{(j)} \in \mathbb{R}^{L \times {\frac{H}{P}} \times {\frac{W}{P}}}$.

\begin{algorithm}[hbt!]
\caption{SPA Training Process}
\label{alg:preference_gmm}
\KwIn{Image $\mathbf{x}$, user-preferred representative segmentation candidate $\mathbf{r_u}$}
\KwOut{Preference-aware Segmentation $\mathbf{y_{app}}$}

\tcc{\textbf{Definitions}}
Preference Distribution Generator: $E_{\theta_{\mu_m;\sigma_m;\pi_m}}(\cdot)$ generating $\{(\mu_m,\sigma_m^{2},\pi_m)\}_{m=1}^{M}$ to construct $p_{\theta}$; Interaction Encoder: $E_{I}(\cdot)$; Number of samples: $N$; Number of representative segmentation candidates: $K$;

\SetKwFunction{FGenMask}{GenMask}
\SetKwProg{Fn}{Function}{:}{\KwRet}
\Fn{\FGenMask{$\mathbf{e_x}$, $p_{\theta}$, ${e_u}$}}{
    Sample $n$ embeddings: $\mathbf{z_n} \sim p_{\theta}$, then \:
    $\mathbf{e_{x_{n}}} = \text{ReLU}(\text{Conv}(\mathbf{e_x} \oplus \mathbf{z_n}))$\;
    $\mathbf{y_n} = \text{Decoder}(\mathbf{e_{x_{n}}}, {e_{u}})$\;
    $\mathbf{y_{app}} = {\mathds{1}_{0.5}} \left( \frac{1}{N} \sum_{n=1}^{N} \mathbf{y_n}  \right)$\;
    \KwRet $\mathbf{y_{app}}, \{\mathbf{y_n}\}_{n=1}^{N}$\;

}

\tcc{\textbf{Training Loop}}
\For{\text{t} = 1 to \text{MAX\_TRAINING\_UPDATES}}{
    Initialize $\mathbf{y}$, $p_{\theta}$, $\mathbf{r_{u}}$ as described in Section \ref{ref:train_update}
    
    \For{\text{j} = 1 to \text{MAX\_USER\_ITERATIONS}}{
        Extract: $\mathbf{e_x} = \text{ViT}(\mathbf{x})$, ${e_u} = E_{I}(\mathbf{r_{u}})$\;
        
        \textbf{Call} $\mathbf{y_{app}^{\text{ori}}}, \_ = \FGenMask(\mathbf{e_x}, p_{\theta}, {e_u})$\;

        \tcc{\textbf{Update Preference Distribution Generator}}
        Update mean and variance:
        $\theta^{'}_{\mu_m; \sigma_m} = \theta_{\mu_m; \sigma_m} + \alpha \nabla_{\theta_{\mu_m}; \sigma_m} \mathcal{L}_{\text{CE}}(\mathbf{y^{ori}_{app}}, \mathbf{y})$;

        Compute GMM responsibilities (E-step): 
        $\pi_m^{gt} = \frac{\pi_m p(y \mid \mu_m, \sigma_m^2)}{\sum_{m=1}^{M} \pi_m p(y \mid \mu_m, \sigma_m^2)}$;
                
        Update GMM weights: 
        $\theta^{'}_{\pi_m} = \theta_{\pi_m} + \alpha \nabla_{\theta_{\pi_m}} \mathcal{L}_{\text{MSE}}(\pi_m^{gt}, \pi_m)$; 
        
        \textbf{Call} $\mathbf{y_{app}^{\text{gmm}}}, \_ = \FGenMask(\mathbf{e_x}, p_{\theta} (\theta^{'}_{\mu_m; \sigma_m}; \theta^{'}_{\pi_m}), {e_u})$\;
        
        \tcc{\textbf{Segmentation Update}}
        Update segmentation and interaction encoders:
        $\theta_{PSeg}^{'} = \theta_{PSeg} + \alpha \nabla_{\theta_{PSeg}} \mathcal{L}_{\text{CE}}(\mathbf{y_{app}^{\text{gmm}}}, \mathbf{y})$, 
        $\theta_{E_{I}}^{'} = \theta_{E_{I}} + \alpha \nabla_{\theta_{E_{I}}} \mathcal{L}_{\text{CE}}(\mathbf{y_{app}^{\text{gmm}}}, \mathbf{y})$\;
        
        \textbf{Call} $\mathbf{y_{app}^{\text{fin}}}, \mathbf{\{y_{n}^{\text{fin}}\}}_{n=1}^{N} = \FGenMask(\mathbf{e_x'}, p_{\theta} (\theta^{'}_{\mu_m; \sigma_m}; \theta^{'}_{\pi_m}), {e_u'})$\;

        \tcc{\textbf{Simulate User Selection for the Next Iteration}}
        Get $K$ Representative Candidates: 
        $\{\mathbf{{r_k}}\}_{k=1}^{K} = \text{KMeans}(\{\mathbf{y_{n}^{\text{fin}}}\}_{n=1}^{N})$\; 
        
        Select preferred representative candidate: 
        $\mathbf{r_u} = \arg\min_{\mathbf{r_k}} \| \mathbf{r_k} - \mathbf{y_{app}^{\text{fin}}} \|_2$;
    }
}
\end{algorithm}

We then employ a mask decoder to predict segmentation masks $\{\mathbf{{y}_n}^{(j)}\}_{n=1}^{N}$  from $\{\mathbf{e_{x_n}}^{(j)}\}_{n=1}^{N}$ and the interaction embedding $e_u^{(j)}$ (Section \ref{section: Space Updating}). The decoder follows the same architecture as SAM \cite{kirillov_segment_2023}. The final preference-aware prediction is aggregated as  $\mathbf{y_{app}}^{(j)} = {\mathds{1}_{0.5}} \left( \frac{1}{N} \sum_{n=1}^{N} \mathbf{y_n}^{(j)} \right)$.

\subsection{Preference Adaption with Human Feedback} \label{section: Space Updating}
\noindent \textbf{Receiving user's choice as preference.} Beyond $\mathbf{y_{app}}^{(j)}$, our model also provides $K$ Representative Segmentation Candidates $\{\mathbf{r_k}^{(j)}\}_{k=1}^{K}$ for further selection, as shown in the lower part of Fig. \ref{fig:Network}. These candidates reflect distinct possible ways to adapt the prediction. To create $\{\mathbf{r_k}^{(j)}\}_{k=1}^{K}$, K-means clustering is applied to the predictions $\{\mathbf{{y}_n}^{(j)}\}_{n=1}^{N}$. Each cluster produces a representative segmentation candidate $\mathbf{r_k}^{(j)} \in \{{\mathbf{r_k}}^{(j)}\}_{k=1}^K$ as the centroid of the cluster. We then highlight the pixel-wise differences $\{\mathbf{\Delta_{r_k}}^{(j)}\}_{k=1}^{K}$ between $\mathbf{{y}_{app}}^{(j)}$ and $\mathbf{r_k}^{(j)}$ to the users. The positive and negative differences are represented in different colors for the convenience of selection. The users then pick their preferred $\mathbf{\Delta_{r_u}}^{(j)}$, which determines the corresponding representative segmentation candidate $\mathbf{r_{u}}^{(j)}$. The $\mathbf{r_{u}}^{(j)}$ is then sent into the interaction encoder mapping into a $L$-dimensional feedback embedding $e_u$ (Fig. \ref{fig:Network}). We use SAM prompt encoder, including its point and mask branches, as our interaction encoder, allowing it to accept both point and mask inputs.

\noindent \textbf{Updating preference distribution.} 
Given the interaction embedding ${e^{(j)}_u}$, the preference distribution $p_{\theta}^{(j)}(z)$ will be updated to $p_{\theta}^{(j+1)}(z)$  with parameters $\theta = \{(\mu^{(j+1)}_m, {\sigma^2_{m}}^{(j+1)}, \pi^{(j+1)}_m)\}_{m=1}^{M}$. Specifically, since obtaining ground-truth embedding samples directly from a given mask is challenging, we replace the analytically updated GMMs with a neural network-based predictor, that we call the Preference Distribution Generator. It adaptively constructs GMMs by predicting their parameters $\theta$ based on human feedback $e^{(j)}_u$. This approach employs amortized inference to efficiently estimate GMM parameters. Given $e^{(j)}_u$, we use six forward layers to jointly predict $\mu^{(j)}_m$ and ${\sigma_{m}}^{(j)}$ from a $2m$-length vector, and use another six forward layers to predict $\pi^{(j)}_m$ (Fig. \ref{fig:Network} lower part). The predictor of $\mu^{(j)}_m$ and ${\sigma_{m}}^{(j)}$ is updated by the backpropagation from the final supervision $y$, while the predictor 
of $\pi^{(j)}_m$ is separately updated by the supervision from ground-truth $\pi^{GT(j)}_m$, since given $\mu^{(j)}_m$ , ${\sigma_{m}}^{(j)}$,  and $\mathbf{y}$, we can get the analytical solution for $\pi^{(j)}_m$  from the GMM E-step (Algorithm \ref{alg:preference_gmm}).

\noindent \textbf{Segmentations with iteratively aligned preference.} In the next iteration, new latent variables $\{\mathbf{z_n}^{(j+1)}\}_{n=1}^{N}$ are sampled from the updated preference distribution $p_{\theta}^{(j+1)}(z)$ and undergo \textit{Preference-aware Segmentation} together with the input image $\mathbf{x}$ and interaction embedding $e_u^{(j)}$.

\noindent \textbf{Inference across input images.} For each input image $\mathbf{x}$, $r_{u}$ is initialized as a user point prompt (similar to SAM), and an initial $p_{\theta}(z)$ is predicted based on $r_{u}$.

\subsection{Training Details} \label{sec:train_details}
In training, for each $\mathbf{x}$, we initialize the ground truth $\mathbf{y}$ as a stochastic combination of collected multi-rater labels: $\mathbf{y} = {\mathds{1}}_{0.5}\left(\sum{A \times Y} \right)$, where $A$ represents the weights uniformly sampled from $[0,1]$  and $Y$ denotes the set of user annotations. We then simulate user interaction and iteratively update the distribution generator to produce \( p_{\theta}^{(j)}(z) \) from \( r_u \), and the segmentation model to predict \( \{\mathbf{y_n}^{(j)}\}_{n=1}^{N} \) based on \( p_{\theta}^{(j)}(z) \). The initial $\mathbf{r_u}$ is simulated by randomly selecting a point from the ground truth. Supervision is then applied to ensure that the final jointly produced aggregation \( \mathbf{y_{app}}^{(j)} \) matches ground truth $\mathbf{y}$. The full training pipeline is in Algorithm \ref{alg:preference_gmm}.

%% file: sec/4_experiment.tex
\begin{table*}[hbt!]
\centering
\caption{\textbf{SPA Outperforms the  SOTA in Dice Score (\%).} We compared deterministic, uncertainty-aware, and interactive models with Dice Score as the metric. SAM-series models use clicks for interaction, while SAM-U uses bounding boxes. SPA, with its multi-choice representative segmentation candidate mechanism, consistently outperforms the other models for the seven tasks. 1-Iter and 3-Iter indicate performance after one and three iterations, respectively.}
\vspace{-5pt}
\resizebox{0.85\linewidth}{!}{
\begin{tabular}{c|c|cc|cccccccc}
\hline
\hline
 Methods   & Category  & 1-Iter & 3-Iter & REFUGE2  & LIDC     & BrainTumor   & Prostate1 & Prostate2 &BrainGrowth                & Kidney   & Ave            \\ \hline
UNet    & \multirow{3}{*}{Det} & \checkmark  &  & 68.94       & 62.99        & 87.30                     & 83.89               & 77.22        & 62.02                          & 82.40       & 74.96                                     \\
TransUNet  & & \checkmark& & 80.83       & 64.09        & 90.14                     & 83.35               & 68.34        & 86.58                          & 52.99       & 75.19                                    \\ 
SwinUNet   & & \checkmark& & 78.67       & 59.45        & 91.23                     & 82.02               & 74.19        & 74.88                          & 69.41       & 75.69                                     \\ 
\hline
\begin{tabular}[c]{@{}c@{}} Ensemble UNet\end{tabular}   & \multirow{5}{*}{Unc} &\checkmark  &
             & 70.75       & 63.84        & 90.56                     & 85.27               & 79.07        & 71.69             & 89.30       & 78.64                                     \\  
ProbUnet   & & \checkmark&  & 68.93       & 48.52        & 89.02                     & 72.13               & 66.84        & 75.59                         & 75.73       & 70.96                                     \\
LS-Unet    &  & \checkmark& & 73.32       & 62.05        & 90.89                    & 87.92               & 81.59        & 85.63                          & 72.31       & 79.10                                     \\ 
MH-Unet   & & \checkmark &   & 72.33       & 62.60        & 86.74                     & 87.03               & 75.61        & 83.54                          & 73.44       & 77.32                                     \\  
MRNet    &  & \checkmark &  & 80.56       & 63.29        & 85.84                    & 87.55               & 70.82        & 84.41                          & 61.30       & 76.25                                     \\ 
\hline
\begin{tabular}[c]{@{}c@{}} SAM\end{tabular}  & \multirow{3}{*}{Int} & \checkmark &
       & 82.59       & 66.68        & 91.55                    & 92.82               & 77.04        & 86.63                          & 85.72       & 83.29                                     \\
MedSAM   & & \checkmark & & 82.34       & 68.42        & 92.67                     & 89.69               & 74.70        & 85.91                          & 78.02      & 81.68                                     \\ 
MSA & & \checkmark & & 83.03       & 66.88        & 88.16                    & 89.06               & 68.94        & 80.62                         & 25.29       & 71.71                                     \\ \hline
SAM-U V1 & \multirow{3}{*}{Unc-Int} &  \checkmark & & 82.45       & 62.24        & 92.67                     & 81.46               & 66.56        & 87.79                          & 89.50       & 80.38                                     \\
SAM-U V2 & &  \checkmark & & 80.66       & 64.82        & 93.11                     & 91.89               & 72.91        & 87.51                          & 90.74       & 83.09                                     \\ 
\rowcolor{cyan!40!white!20} 
\textbf{SPA}&   &  \checkmark &    & \textbf{83.47}       & \textbf{88.07}        & \textbf{94.29}                    & \emph{\textbf{93.12}}               & \textbf{83.34}        & \textbf{88.14}                          & \textbf{94.08}       & \textbf{89.22}                                     \\ \hline
\begin{tabular}[c]{@{}c@{}} SAM\end{tabular} & \multirow{3}{*}{Int} &   & \checkmark
       & 82.61       & 66.71        & 92.14                     & 92.72               & 77.54        & 86.58                          & 90.43       & 84.10                                     \\ 
MedSAM   &   & & \checkmark  & 82.13       & 68.45        & 93.26                   & 90.05               &  73.81        & 86.09                          & 79.88       & 81.95                                  \\ 
MSA&   & & \checkmark & 83.08       & 66.87        & 91.25                  & 90.22               & 71.34        & 81.87                          & 46.76       & 75.91                                     \\ \hline
SAM-U V1 &   \multirow{3}{*}{Unc-Int} & & \checkmark  & 82.10       & 62.84        & 92.31                     & 81.79               & 66.74        & 87.84                          & 89.24       & 80.40                                     \\ 
SAM-U V2 &   & & \checkmark & 80.54       & 65.44        & 92.40                     & 90.00               & 73.17        & 87.87                          & 91.35       & 82.96                                     \\ 
\rowcolor{cyan!40!white!20}
\textbf{SPA}&    & & \checkmark       & \emph{\textbf{85.42}}       & \emph{\textbf{88.56}}        & \emph{\textbf{94.31}}                   & \textbf{92.97}               & \emph{\textbf{84.05}}        & \emph{\textbf{88.18}}                          & \emph{\textbf{94.26}}       & \emph{\textbf{89.68}}                                     \\ \hline \hline

\end{tabular}
}\label{tab:SOTA result}
\vspace{-10pt}
\end{table*}

\section{Experiment}
We conducted experiments to validate the effectiveness of SPA across seven uncertainty segmentation tasks represented by multi-clinician, using data from three medical imaging modalities: color fundus images, lung lesion and kidney CT scans, brain and prostate MRI scans. SPA consistently achieves SOTA performance compared to deterministic, uncertainty-aware, and interactive models. Notably, SPA outperforms interactive models with significantly fewer iterations and it demonstrates strong generalization on unseen users. Human evaluation further shows that SPA requires less time and fewer interactions to meet human expectations. Our results also show that predictions increasingly align with selected users while diverging from those excluded. Ablation studies highlight the necessity of each component for optimal segmentation. Additional analysis of representative segmentation candidates' similarities across iterations and prediction changes after each interaction are detailed in the appendix.

\subsection{Dataset}
\textbf{REFUGE2} benchmark \cite{fang_refuge2_2022} is a publicly available fundus image dataset for glaucoma analysis, including optic cup segmentation. REFUGE2 includes annotations from seven ophthalmologists, each with an average of eight years of experience. It contains 400 training and 400 test images.

\noindent
\textbf{LIDC-IDRI} benchmark \cite{armato_lung_2011, clark_cancer_2013} originally consisted of 3D lung CT scans with semantic segmentations of possible lung abnormalities, annotated by four radiologists. We use a pre-processed version from \cite{kohl_hierarchical_2019} of 15,096 2D CT images. After an 80-20 train-test split, the training and test datasets contain 12,077 and 3,019 images, respectively.

\noindent
\textbf{QUBIQ} benchmark \cite{li_qubiq_2024} consists of one MRI brain tumor task (three annotations, 28 training cases, 4 testing cases); two MRI prostate tasks (six annotations, 48 training cases, 7 testing cases); one MRI brain-growth task (seven annotations, 34 training cases, 5 testing cases); and one CT kidney task (three annotations, 20 training cases, 4 testing cases).

\subsection{Implementation Details}\label{sec:implementation}
Our network was implemented using PyTorch v1.12 and trained/tested on an RTX A4000 with 16GB of memory. During training, we used the Adam optimizer with an initial learning rate of $1e^{-4}$ and adjusted it using StepLR strategy. $L=256$ features were extracted to generate feature embeddings and $N = 48$ latent variables were sampled from the preference distribution $p^{(j)}_{\theta}(z)$, which was modeled with $M = 16$ Gaussian components. We generated $K = 4$ representative segmentation candidates to allow multi-choice selection and set $MAX\_USER\_{ITERATIONS}$ to 6. It takes 0.2 second per iteration to generate the prediction and the representative segmentation candidates, acceptable for human-in-the-loop settings.

We used \textbf{SAM-ViT/B} as the segmentation backbone. Additional implementation details are provided in the appendix. Deterministic segmentation methods with multiple annotations were trained with majority vote labels. For SAM-series interactive models, click or bounding box prompts were uniformly generated based on the original model settings.


\subsection{Experimental Results}
\subsubsection{Performance Comparison with SOTA Methods}
To demonstrate the advantages of SPA, we compared it with SOTA methods, classified into deterministic methods (UNet \cite{ronneberger_u-net_2015}, TransUNet \cite{chen_transunet_2021}, SwinUNet \cite{cao_swin-unet_2021}), uncertainty-aware methods (Ensemble UNet, ProbUnet \cite{kohl_probabilistic_2018}, LS-Unet \cite{jensen_improving_2019}, MH-Unet \cite{guan_who_2018}, MRNet \cite{ji_learning_2021}), and interactive methods (SAM \cite{kirillov_segment_2023}, MedSAM \cite{ma_segment_2024}, MSA \cite{wu_medical_2023}). We also compared SPA with SAM-U \cite{deng_sam-u_2023}, an uncertainty-interactive method that simply introduces uncertainty by generating multiple prompts. SAM-U was evaluated using both SAM and MedSAM backbones with bounding boxes as the interaction strategy, named SAM-U V1 and SAM-U V2, respectively. Other SAM-series methods relied on user clicks for interactions. Results for interactive models were reported after one and three iterations.

\begin{figure*}[hbt!]
    \centering
    \includegraphics[scale=0.61]{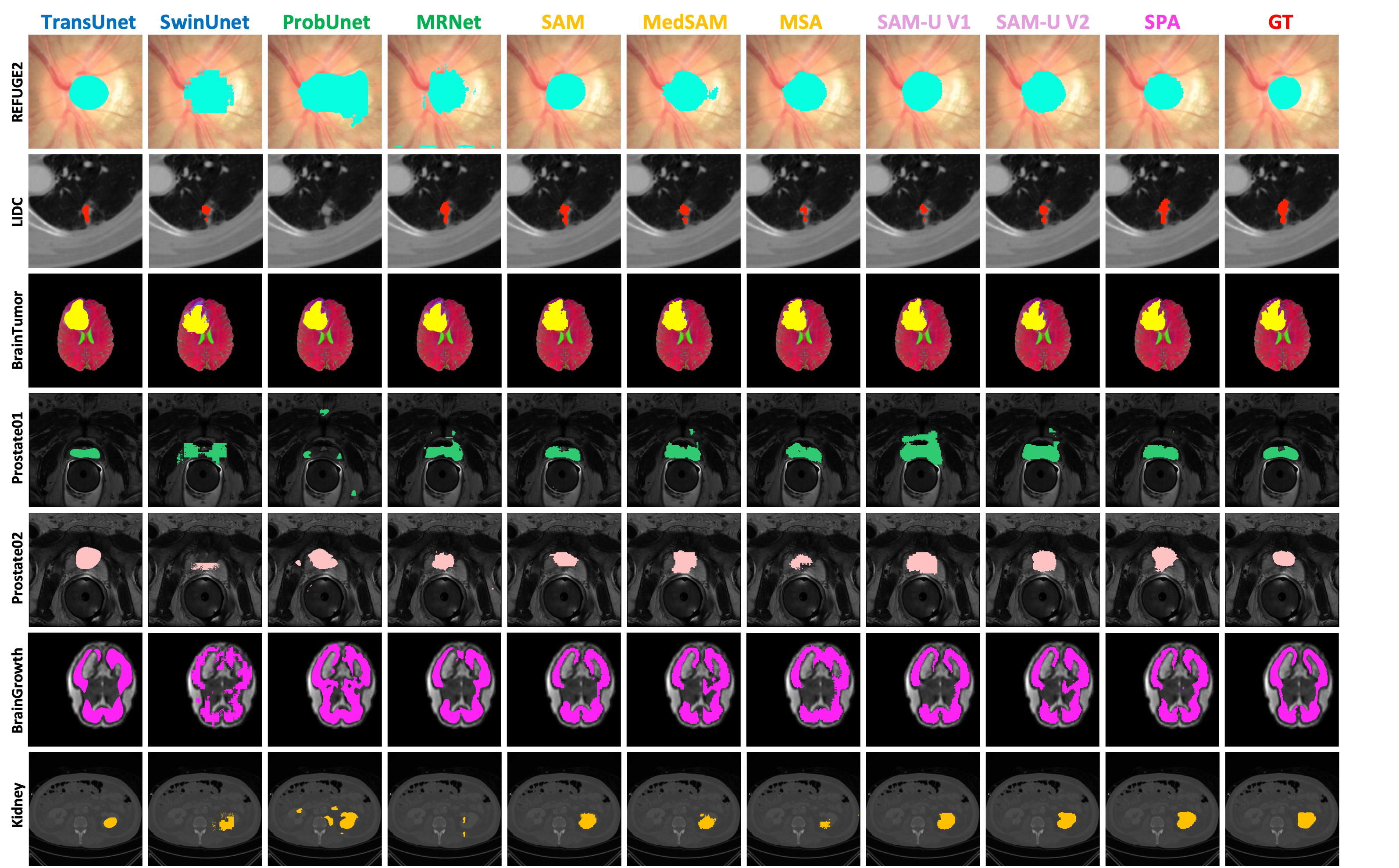} 
    \vspace{-5pt}
    \caption{\textbf{SPA Shows Superior Segmentation Visualization.} Visual comparison of segmentation results with deterministic, uncertainty-aware, and interactive models after six iterations. SPA provides better adaptability, particularly at boundary regions.}
    \label{fig:Visualisation}
\end{figure*}

\noindent \textbf{SPA consistently outperforms all methods}, achieving an average Dice Score of 89.68\% after three iterations. Table \ref{tab:SOTA result} provides a quantitative comparison using Dice Score as the metric. The improvement is particularly notable for the LIDC segmentation task, where SPA improves on current SOTA methods by 20\%. Even after one iteration, SPA demonstrates superior performance relative to the other methods. Fig. \ref{fig:Visualisation} shows visual comparisons between SPA and other SOTA methods, presenting segmentations after six iterations for the interactive models. The results qualitatively indicate that the segmentations predicted by SPA align more closely with the ground truth, especially in the boundary regions. 

\begin{figure*}[hbt!]
    \centering
    \vspace{-5pt}
    \includegraphics[scale=0.52]{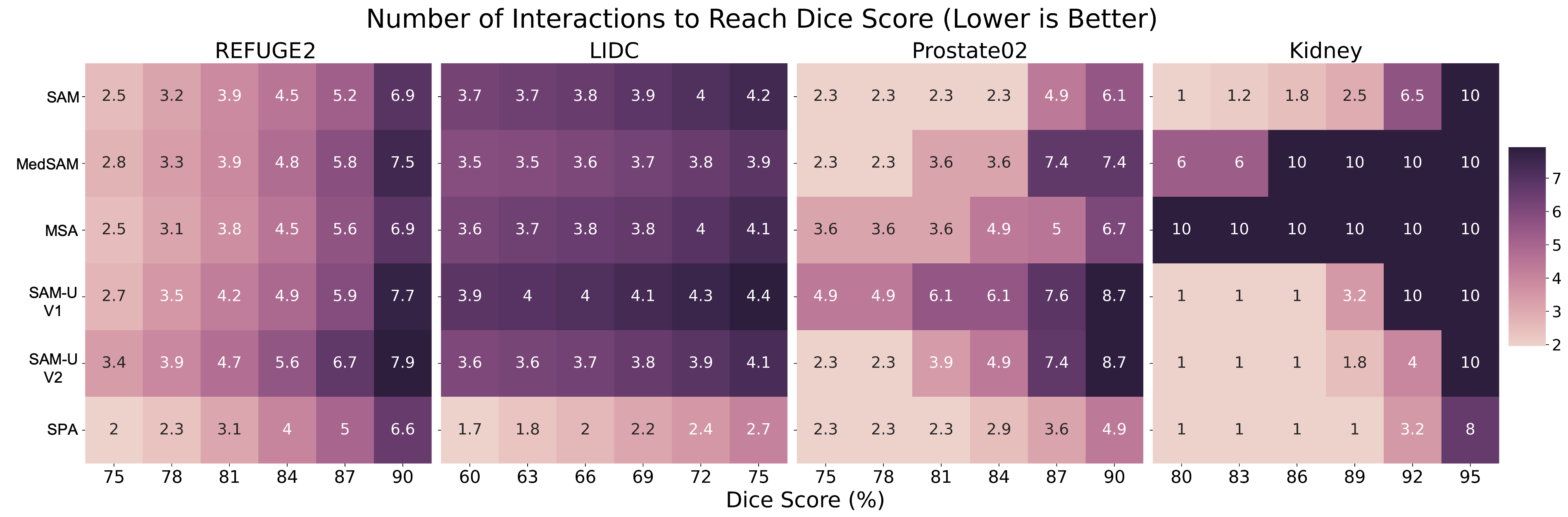} 
    \vspace{-10pt}
    \caption{\textbf{SPA Demonstrates Extraordinary Efficiency.} Efficiency analysis comparing the average number of iterations required to reach specific Dice Scores across interactive models. Models that failed to reach the target Dice Score within six iterations are assigned an iteration count of ten. Failure rate is provided in the appendix. SPA consistently achieves lower failure rates and requires fewer iterations.}
    \label{fig:converge}
\vspace{-10pt}
\end{figure*}


\subsubsection{Efficiency Analysis on Different Interactive Models}
We conducted an efficiency analysis to quantify the number of iterations required to reach specific Dice Scores, with a maximum of six iterations. Models that failed to reach the target Dice Score within the limit were assigned an iteration count of ten. The failure rate is provided in the appendix. Our proposed method, SPA, consistently has lower failure rates and requires fewer iterations to achieve superior segmentation performance compared to other models.

\noindent \textbf{SPA outperforms all other interactive models in terms of interaction efficiency}, as shown in Fig. \ref{fig:converge}. On the REFUGE2 dataset, it requires fewer than five iterations to achieve a Dice Score of 87\%, whereas SAM-U V2 needs 5.6 iterations to reach a Dice Score of just 84\%. In the LIDC dataset, SPA achieves 75\% Dice Score with only 2.7 iterations on average, while other models typically require four or more iterations. Similarly, on the Kidney dataset, SPA reaches a Dice Score of 92\% with 3.2 iterations, while other SAM-series models often need four to ten iterations. Notably, SPA reaches a Dice Score of 95\% within eight iterations, but other models struggle to converge within the ten-iteration limit.


\begin{table*}[hbt!]
\centering
\caption{\textbf{SPA Demonstrates Strong Prediction Alignment with Individual Users.} Alignment analysis by Dice Score (\%) on the REFUGE2 optic cup segmentation task. The ground truth combines annotations from {\textcolor{red}{Clinicians 2, 4, 5, 6, and 7}}, excluding {\textcolor{blue}{Clinicians 1 and 3}}. Predictions are compared with individual clinicians (columns) across interaction iterations (rows). The predictions show increasing alignment with included clinicians and divergence from excluded clinicians.}
\vspace{-5pt}
\resizebox{1\linewidth}{!}{%
\begin{tabular}{c|ccccccc}
\hline \hline
Iteration &  \textcolor{blue}{Clinician 1}     & \textcolor{red}{Clinician 2}     & \textcolor{blue}{Clinician 3}    & \textcolor{red}{Clinician 4}     & \textcolor{red}{Clinician 5}     & \textcolor{red}{Clinician 6}     & \textcolor{red}{Clinician 7}                  \\ \hline
1 & 75.02           & 78.92           & 83.98           & 78.25           & 80.81           & 84.86           & 59.35                      \\
2 & 74.03 (-1.32\%) & 80.58 (+2.10\%) & 84.03 (+0.05\%) & 78.88 (+0.81\%) & 81.92 (+1.37\%) & 85.62 (+0.90\%) & 60.43 (+1.82\%)  \\
3 & 73.61 (-1.88\%) & 80.88 (+2.48\%) & 83.68 (-0.36\%) & 79.16 (+1.16\%) & 82.17 (+1.68\%) & 85.87 (+1.19\%) & 60.79 (+2.43\%)  \\
4 & 73.35 (-2.23\%) & 81.29 (+3.00\%) & 83.40 (-0.69\%) & 79.19 (+1.20\%) & 82.32 (+1.87\%) & \textbf{86.06 (+1.41\%)} & 61.06 (+2.88\%)  \\
5 & 73.27 (-2.33\%) & 81.35 (+3.08\%) & 83.32 (-0.79\%) & 79.17 (+1.18\%) & 82.48 (+2.07\%) & 85.90 (+1.23\%) & 61.14 (+3.02\%)  \\
6 & \textcolor{blue}{\textbf{73.15 (-2.49\%)}} & \textcolor{red}{\textbf{81.54 (+3.32\%)}} & \textcolor{blue}{\textbf{83.30 (-0.81\%)}} & \textcolor{red}{\textbf{79.23 (+1.25\%)}} & \textcolor{red}{\textbf{82.65 (+2.28\%)}} & \textcolor{red}{86.03 (+1.38\%)} & \textcolor{red}{\textbf{61.34 (+3.35\%)}} 
\\ \hline\hline
\end{tabular}
}\label{tab:different_clinician}
\vspace{-10pt}
\end{table*}

\noindent \textbf{SPA shows generalization and robustness to unseen preferences}. To show this, we conducted a leave-one-user-out experiment three times, each time leaving a different user out during the training phase. The reported results are averaged across these three runs, while individual user statistics are provided in the appendix. On REFUGE2 optic cup segmentation task, SPA achieves a 75\% Dice Score in an average of 3.01 iterations, whereas other models require more than 3.6 iterations to reach similar accuracy. For a Dice Score of 84\%, SPA requires an average of 5.02 iterations, surpassing the second-best model, MedSAM, by 0.94 iterations. In addition, SPA is transferable between images for the same user. The Dice Score improves from 91.68\% to 92.43\% on Kidney dataset by using user-adapted GMM parameters.

\noindent \textbf{SPA's improved efficiency is further substantiated with a human user study}. We invited five medical professionals, each with over five years of graduate-level expertise, to annotate 100 fundus images from the REFUGE2 dataset. Across all annotators, SPA consistently required fewer interactions and less time compared to MedSAM. For example, one individual required an average of 6.62 seconds and 4.33 iterations per image with MedSAM, but only 4.34 seconds and 2.46 iterations with SPA. Another annotator took 7.40 seconds and 5.60 iterations per image with MedSAM, but just 4.77 seconds and 3.58 iterations with SPA. Similarly, for the remaining annotators, SPA required an average of 4.56, 5.03, and 4.82 seconds per image, compared to 6.91, 8.04, and 7.60 seconds with MedSAM. This significant reduction in both time and interactions highlights SPA's superior efficiency. User study details are provided in the appendix.


\subsubsection{Prediction Alignment with Clinician Feedback in Model Adaptation}
Table \ref{tab:different_clinician} shows prediction alignment results for optic cup segmentation on the REFUGE2 dataset, comparing SPA's predictions with individual clinicians' annotations after each iteration. In this case, the ground truth is defined as the weighted average of annotations from Clinicians 2, 4, 5, 6, and 7, excluding Clinicians 1 and 3. After each interaction, the Dice Scores for the included clinicians consistently improve, indicating that SPA is adapting to the desired context. For example, Clinician 2's score increases from 78.92\% to 81.54\% (+3.32\%), with similar positive trends for Clinicians 4, 5, 6, and 7 (+ 1.25\%, +2.28\%, +1.38\%, and +3.35\%, respectively). In contrast, the Dice Scores for the excluded clinicians, such as Clinicians 1 and 3, show a consistent decline (-2.49\% and -0.81\%), indicating that the model is effectively excluding irrelevant clinicians. This multi-choice-based refinement demonstrates SPA's ability to align predictions with the included clinicians while excluding those not part of the ground truth. Visualization of this alignment is provided in the appendix.

\subsubsection{Ablation Study}
In this section, we conducted an ablation study on key components of SPA, including random sampling from the preference distribution (Random Gaussian), updating the mean and variance (Gaussian), and adjusting distribution weights (Mixture Gaussian). Table \ref{tab:ablation} shows the ablation results, with segmentation performance evaluated by Dice Score on the REFUGE2 and Kidney datasets after three iterations.

When only randomly sampling is used without calibrating the mean, variance, or weight of the preference distribution, the Dice Scores are 80.29\% for REFUGE2 and 90.05\% for the Kidney dataset. Training the mean and variance improves the scores to 84.12\% and 92.06\%, respectively. Additionally, training the distribution weights alone raises the scores to 82.87\% for REFUGE2 and 92.29\% for Kidney. Finally, combining all three components which results to calibrating distribution mean, variance, and weight to form the preference distribution yields the highest performance, with Dice Scores of 85.42\% for REFUGE2 and 94.26\% for the Kidney dataset. This highlights the complementary benefits of each component in achieving optimal segmentation performance.

\begin{table}[hbt!]
\centering
\vspace{-5pt}
\caption{\textbf{Effectiveness of Network Modules in SPA.} Ablation study evaluating the impact of network components on segmentation performance for the REFUGE2 and Kidney datasets after three iterations. The table compares random sampling (Random Gaussian), updating the mean and variance (Gaussian), and updating distribution weights (Mixture Gaussian). Combining all modules achieves the highest Dice Scores, highlighting their complementary benefits in optimizing segmentation performance.}
\vspace{-5pt}
\resizebox{1\linewidth}{!}{%
\begin{tabular}{ccccc}
\hline\hline
Random  Gaussian & Gaussian & Mixture Gaussian & REFUGE2 & Kidney \\ \hline
\checkmark &   &   &   80.29 & 90.05  \\
\checkmark & \checkmark  &          &  84.12 & 92.06  \\
\checkmark &           & \checkmark &  82.87 & 92.29\\
\rowcolor{cyan!40!white!20}
\checkmark &    \checkmark   & \checkmark & \textbf{85.42} &  \textbf{94.26} \\ \hline \hline
\end{tabular}%
}\label{tab:ablation}
\vspace{-15pt}
\end{table}

%% file: sec/5_conclusion.tex
\section{Conclusion}
In this work, we introduce SPA, a novel segmentation framework that efficiently adapts to user preferences with minimal human effort. By offering users multi-choice options based on image uncertainties at interactions, SPA reduces user workload and ensures preference-specific predictions. The proposed preference distribution allows the model to dynamically adapt to user feedback during inference, accelerating convergence and enhancing interaction efficiency. Reported experiments show that SPA outperforms deterministic, uncertainty-aware, and interactive SOTA models, demonstrating strong adaptability in different clinical contexts while requiring significantly less time and effort. These results highlight SPA's potential to contribute to improving clinical workflows in real-world medical applications.

\section*{Acknowledgement}
Jiayuan Zhu is supported by the Engineering and Physical Sciences Research Council (EPSRC) under grant EP/S024093/1 and Global Health R\&D of Merck Healthcare, Ares Trading S.A. (an affiliate of Merck KGaA, Darmstadt, Germany), Eysins, Switzerland (Crossref Funder ID: 10.13039/100009945). Junde Wu is supported by the Engineering and Physical Sciences Research Council (EPSRC) under grant EP/S024093/1 and GE HealthCare. Cheng Ouyang is supported by UKRI grant EP/X040186/1.

%% file: sec/X_suppl.tex
\setcounter{page}{1}
\maketitlesupplementary

\section{Motivation Details} \label{app:detailed_motivation}
\begin{table}[hbt!]
\centering
\vspace{-10pt}
\caption{A preliminary experiment testing the impact of individual clinicians, conducted for the optic cup segmentation on REFUGE2 test set under U-Net's structure with Dice Score (\%). The results indicate that the segmentation performance is consistent for individual clinician but varies significantly across different clinicians.}
\vspace{5pt}
\resizebox{1\linewidth}{!}{%
\begin{tabular}{c|cccc}
\hline
\hline
        & Clinician 1    & Clinician 2    & Clinician 3    & Clinician 4    \\ \hline
Model 1 & \textbf{71.28} & 60.28          & 47.67          & 52.25          \\
Model 2 & 61.06          & \textbf{66.46} & 63.30          & 63.84          \\
Model 3 & 52.30          & 62.29          & \textbf{69.30} & 64.06          \\
Model 4 & 53.05          & 62.72          & 65.45          & \textbf{67.19} \\ \hline\hline
\end{tabular}%
}\label{tab:detailed_motivation}
\end{table}

Medical image uncertainty is often reflected in the varying user preferences, causing inconsistent annotations between different clinicians. To further explore this, we conducted a preliminary experiment to quantitatively demonstrate that individual clinicians exhibit consistent segmentation patterns, while significant variation exists between different clinicians. In this experiment, we trained U-Net models \cite{ronneberger_u-net_2015} using each clinician's annotations for optic cup segmentation with a subset of the REFUGE2 dataset. This resulted in four distinct models (Model 1-4), each corresponding to a different clinician (Clinician 1-4). Table \ref{tab:detailed_motivation} shows the segmentation performance of each model when evaluated against different clinicians. Notably, each model performed best when trained and tested with the same clinician's annotations, but its performance dropped significantly when evaluated against other clinician's annotations. 

These observations suggest that each clinician exhibits a distinct and consistent annotation pattern, which directly influences the performance of the segmentation model. For example, Model 1 achieved its best performance when tested on Clinician 1's annotations, with a Dice Score of 71.28\%. However, its performance decreased substantially when evaluated against the annotations of Clinician 2, 3, or 4, with scores as low as 47.67\%. This trend persisted across all models, indicating that segmentation performance declines significantly when a model is trained on one clinician's annotations and tested on another's. 

As each clinician's annotation behavior is informed by their specific preference, this finding suggests that adapting to these preference-driven behaviors could improve preference-specific segmentation predictions. We further hypothesize that not only are annotation patterns consistent within individual clinicians, but their interaction behaviors during interactive segmentation are also likely to be consistent. This hypothesis motivated the development of the SPA model, which is designed to adaptively learn and adjust to each clinician’s specific preference through human interactions. By dynamically incorporating clinician feedback, SPA refines the segmentation process in response to individual interactions, aligning the model’s predictions with the individual clinician preference.

\begin{figure*}[ht]
    \begin{minipage}{0.4\linewidth}
            \centering
            \includegraphics[scale=0.3]{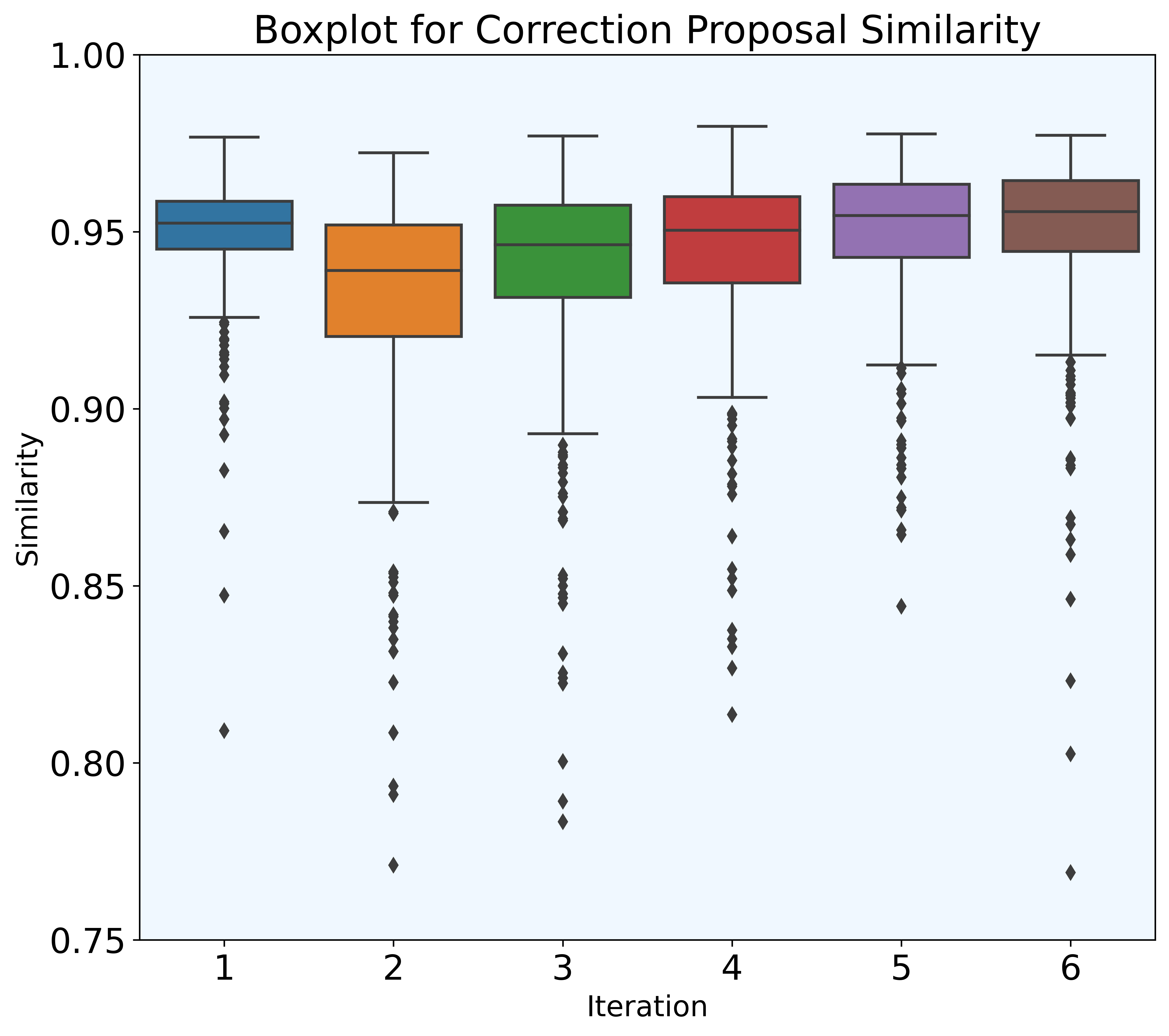} 
    \end{minipage}\,
\begin{minipage}{0.6\linewidth}
        \centering
        \includegraphics[scale=0.2]{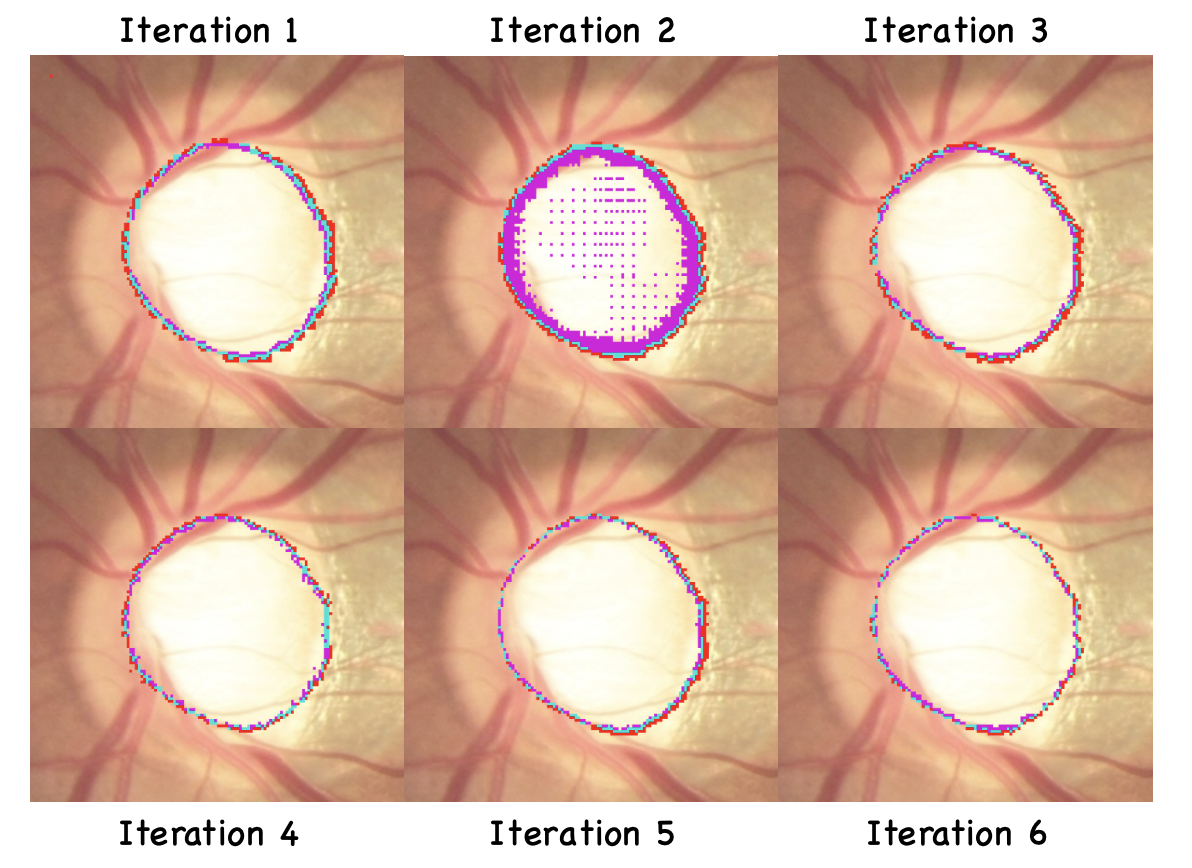} 
    \end{minipage}
\caption{\textbf{Representative Segmentation Candidates Converge Over Iterations.} (a) Boxplot illustrating the similarity (measured by Dice Score) between representative segmentation candidates across multiple iterations. (b) Visualization showing the disagreement among four correction candidates at each iteration. Red regions indicate areas recognized by only one candidate as part of the target, light blue by two candidates, and purple by three candidates. As iterations progress, areas of disagreement shrink and alignment increases, reflecting the model's adaptation to user interactions and the reduction of uncertainty over time, especially from iterations 2 to 6.}\label{fig:proposal_similarity}
\end{figure*}


\section{Theoretical Proof}
\label{app: Theoretical Proof}
Let $D = \{\mathbf{{r_u}}^{(j)}\}_{j=1}^{J}$ represent $J$ interactions from user $u$, where each interaction $\mathbf{{r_u}}^{(j)}$ is generated i.i.d. from a specific component $\mathcal{N}(\mu_u, \sigma_{u}^2)$. 
The posterior probability that the samples (interactions) $D$ comes from user (Gaussian component) $p$ is given by:
\begin{equation}
\label{eq:1}
  P(U = u \mid D) = \frac{\pi_u \prod_{j=1}^{J} N(\mathbf{{r_u}}^{(j)} \mid \mu_u, \sigma_u^2)}{\sum_{m=1}^{M} \pi_m \prod_{j=1}^{J} N(\mathbf{{r_u}}^{(j)} \mid \mu_m, \sigma_m^2)}
\end{equation}
The average log-likelihood of each interaction belonging to user $u$ is:
\begin{equation}
\label{eq:3}
\frac{1}{J}\log L_u(D) = \frac{1}{J}\log \pi_u + \frac{1}{J}\sum_{j=1}^{J} \log N(\mathbf{{r_u}}^{(j)} \mid \mu_u, \sigma_u^2)
\end{equation}
As $J \to \infty$, the empirical average converges to the expected value under the true distribution $\mathcal{N}(\mu_u, \sigma_{u}^2)$:
\begin{equation}
\resizebox{0.91\linewidth}{!}{
$\lim_{J \to \infty} \frac{1}{J} \sum_{j=1}^{J} \log N(\mathbf{{r_u}}^{(j)} \mid \mu_u, \sigma_u^2) = \mathbb{E}_{\mathbf{{r_u}} \sim N(\mu_u, \sigma_u^2)} \left[ \log N(\mathbf{{r_u}} \mid \mu_u, \sigma_u^2) \right]$}
\end{equation}
The expected log-likelihood difference between user $u$ and any other user $i \neq u$ is expressed as the negative Kullback-Leibler (KL) divergence: 
\begin{equation}
\label{eq:5}
\resizebox{0.91\linewidth}{!}{$
\mathbb{E}_{\mathbf{{r_u}} \sim N(\mu_u, \sigma_u^2)} \left[ \log N(\mathbf{{r_u}} \mid \mu_u, \sigma_u^2) - \log N(\mathbf{\Delta_{r_u}} \mid \mu_i, \sigma_i^2) \right] = - D_{\mathrm{KL}} \left( N(\mu_u, \sigma_u^2) \,\|\, N(\mu_i, \sigma_i^2) \right)$}
\end{equation} 
Combining this with Equation \ref{eq:3}, we obtain the likelihood ratio between user $u$ and any other user $i \neq u$:
\begin{equation}
\frac{L_u(D)}{L_i(D)} = e^{-Q D_{\mathrm{KL}}(N(\mu_u, \sigma_u^2) \,\|\, N(\mu_i, \sigma_i^2))} \cdot \frac{\pi_u}{\pi_i}
\end{equation}
Thus, the posterior distribution from Equation \ref{eq:1} for user $u$ becomes:
\begin{equation}
\resizebox{0.91\linewidth}{!}{$
  P(U = u \mid D) = \left[ 1 + \sum_{i \ne u} \left( \frac{\pi_u}{\pi_i} e^{-Q D_{\mathrm{KL}}(N(\mu_u, \sigma_u^2) \,\|\, N(\mu_i, \sigma_i^2))} \right) \right]^{-1}$}
\end{equation}
Since $\mu_i \ne \mu_u$ or $\sigma_i^2 \ne \sigma_u^2$ for $i \neq u$, we have $ D_{\mathrm{KL}}(N(\mu_u, \sigma_u^2) \,\|\, N(\mu_i, \sigma_i^2)) > 0$ \footnote{KL divergence property: $D_{\mathrm{KL}}(A \,\|\, B) \geq 0, D_{\mathrm{KL}}(A \,\|\, B) = 0 \text{ iff } A = B$}. As $J \to \infty$, $e^{-J D_{\mathrm{KL}}} \to 0$ for each $i \ne u$, leading to $P(U = u \mid D) \to 1$. In other words, as $J \to \infty$, $p_\theta(z) \to N(\mu_u, \sigma_u^2)$. Therefore, the preference distribution $p_{\theta}(z)$ can adapt to specific user preferences based on new interactions. 
Since there is no closed-form solution for updating $\theta = \{(\mu_m, \sigma_{m}^2, \pi_m)\}_{m=1}^{M}$, we use MLP blocks with six forward layers and ReLU activations to adjust these parameters, refining the preference distribution $p_{\theta}(z)$ based on interactions. The preference distribution $p_{\theta}(z)$ thus adapts effectively to individual user preferences through interactions, enabling a personalized segmentation model that aligns closely with diverse clinical contexts and user expectations.


\section{Implementation Details} 
\label{app: Ground Truth Definition}
To capture image uncertainty, we generate $N = 48$ predictions by sampling from the preference distribution $p^{(j)}_{\theta}(z)$. 48 predictions is experimentally the best to balance model performance and computational cost. The Dice Score for generating 24, 36, 48 predictions on REFUGE2 after six iterations are 83.05\%, 84.70\% and 86.22\%. Therefore, we choose to sample $48$ predictions as the hyperparameter. Additionally, we generate $K = 4$ representative segmentation candidates to allow users to make a multiple-choice selection. This hyperparameter is set to 4 because it is commonly used in high-stakes tests and is practical in various scenarios \cite{dehnad2014comparison}. As a result, $4$ is chosen as the number of representative segmentation candidates in our medical setting. In addition, we sample the most representative point from the representative segmentation candidate to prevent overfitting, although the segmentation candidates can be directly used. The initial interaction is uniformly selected from the area agreed upon by all annotators. It mimics real-world user behavior, where users first identify a rough target or shape and then refine the boundaries.

In this work, we face the challenge of medical image uncertainty, particularly when using multi-user annotated datasets such as REFUGE2, LIDC-IDRI, and QUBIQ. Each user's annotation reflects their unique interpretations, leading to inherent uncertainty in segmentations. To establish a reliable ground truth for evaluating our SPA method, we adopt a strategy that combines annotations from multiple users. This approach allows us to capture the medical image uncertainty associated with varying human preferences while minimizing potential biases. It ensures model robustness and generalisability.

The use of multi-user datasets is central to modeling medical image uncertainty. While individual user annotations are valuable, they can introduce personal biases or errors and may not fully represent the range of clinical contexts or human preferences. By combining multiple user annotations, we are able to account for the diverse preferences that contribute to uncertainty in medical images. This combination reflects a more balanced and representative ground truth, reducing the impact of any single user’s subjective interpretation. Additionally, when only a limited number of users (e.g., four in the LIDC dataset) are available, it becomes crucial to use their combined annotations to better simulate the diversity of clinical contexts. By integrating multiple user annotations, we more accurately represent shifts in annotation conventions and clinical decision-making processes that occur across different medical environments, thereby capturing the uncertainty inherent in medical image analysis.

During the testing stage, for each image, we randomly generated combinations of user annotations from the multi-user annotated datasets. When four users provided annotations, all possible combinations (individual, pairs, triplets, and full set) were considered, resulting in 15 combinations (4 individual, 6 pairs, 4 triplets, and 1 full set). These combinations were uniformly selected, ensuring equal representation of all potential groupings. The selected annotations were then fused to create a consensus segmentation. This fusion involved averaging the chosen annotations to generate a probability map, which was subsequently binarized to form the final segmentation mask. By incorporating multiple user perspectives, this fused binary segmentation reflects the inherent uncertainty in the data and was used as the ground truth for evaluating the SPA model.

\section{Efficiency Analysis on Different Interactive Models Details}
Models that failed to reach the target Dice Score within the limit were assigned an iteration count of ten. We provide the failure rate statistics for REFUGE2 dataset reaching Dice 70\% and 80\%, for LIDC reaching Dice 60\% and 70\% in Table \ref{tab:failure_rate}. It shows that our method, SPA, consistently achieves fewer failure cases across datasets and thresholds. 

\begin{table}[hbt!]
\caption{Failure rate for REFUGE2 reaching Dice 70\% and 80\%, for LIDC reaching Dice 60\% and 70\%.}
\vspace{-5pt}
\centering
\resizebox{1\linewidth}{!}{%
\begin{tabular}{c|cc|cc|cc|cc|cc|cc}
        & \multicolumn{2}{c|}{SAM} & \multicolumn{2}{c|}{MedSAM} & \multicolumn{2}{c|}{MSA} & \multicolumn{2}{c|}{SAM-U V1} & \multicolumn{2}{c|}{SAM-U V2} & \multicolumn{2}{c}{SPA} \\ \hline
REFUGE2 & 10.2\%      & 28.2\%     & 11.7\%       & 29.2\%       & 9.0\%       & 27.2\%     & 10.5\%        & 33.0\%        & 15.0\%        & 38.7\%        & \textbf{4.7\%}      & \textbf{17.0\%}     \\
LIDC    & 29.4\%      & 32.5\%     & 27.2\%       & 29.8\%       & 29.1\%      & 31.9\%     & 31.9\%        & 34.9\%        & 27.9\%        & 31.3\%        & \textbf{7.0\%}      & \textbf{13.0\%}   
\end{tabular}
}\label{tab:fail}
\vspace{-10pt}
\end{table}\label{tab:failure_rate}

In addition, to further verify the generalization and robustness, we further demonstrate the number of iterations required to reach specific Dice Score for different unseen annotators in Table \ref{tab:adapt} . It shows that our approach, SPA, consistently requires less iterations than other interactive models to reach specific Dice Score on the REFUGE2 dataset, regardless of the annotator.

\begin{table}[hbt!]
\caption{Number of iterations required to reach Dice 75\% and 84\% toward different unseen annotators' (A, B, C).}
\vspace{-5pt}
\centering
\resizebox{1\linewidth}{!}{%
\begin{tabular}{l|ll|ll|ll|ll|ll|ll}
\multicolumn{1}{c|}{}  & \multicolumn{2}{c|}{SAM}                             & \multicolumn{2}{c|}{MedSAM}                          & \multicolumn{2}{c|}{MSA}                             & \multicolumn{2}{c|}{SAM-U V1}                        & \multicolumn{2}{c|}{SAM-U V2}                        & \multicolumn{2}{c}{SPA}                                               \\ \hline
A                      & 6.85                     & 8.70                      & 7.39                     & 9.31                      & 6.78                     & 8.76                      & 6.90                     & 8.64                      & 6.25                     & 8.33                      & \textbf{5.64}                     & \textbf{7.69}                     \\
\multicolumn{1}{c|}{B} & \multicolumn{1}{c}{2.76} & \multicolumn{1}{c|}{5.87} & \multicolumn{1}{c}{2.09} & \multicolumn{1}{c|}{5.03} & \multicolumn{1}{c}{2.49} & \multicolumn{1}{c|}{5.57} & \multicolumn{1}{c}{2.70} & \multicolumn{1}{c|}{5.42} & \multicolumn{1}{c}{3.22} & \multicolumn{1}{c|}{6.79} & \multicolumn{1}{c}{\textbf{1.89}} & \multicolumn{1}{c}{\textbf{4.38}} \\
C                      &     1.72                     &         4.08                  & 1.66                     & 3.58                      & 1.56                     & 3.64                      &    1.66                      &   3.99                        &    1.53                      &           3.57                & \textbf{1.50}                     & \textbf{3.01}                    
\end{tabular}
}\label{tab:adapt}
\vspace{-10pt}
\end{table}

\section{Representative Segmentation Candidate Similarity across Interactions} 
\label{app:Correction Proposal Similarity across Interactions}

In this section, we explore how the similarity between representative segmentation candidates evolves over multiple iterations of human interaction. Our multi-choice approach generates distinct segmentation candidates after each iteration, enabling users to guide the refinement process by selecting the segmentation candidate that they find the most appropriate based on their preference. This analysis investigates how these representative segmentation candidates change as the model adapts to user feedback over time. For illustration, we use the optic cup segmentation task from the REFUGE2 dataset. To facilitate comparison, we directly used the raw K-means clustering results applied to the $N$ predictions $\{\hat{y}_n^{(j)}\}_{n=1}^{N}$. This approach allows us to evaluate the inherent divergence or convergence of the candidates without any post-processing adjustments. To quantify the similarity between candidates, we computed the Dice Score as the similarity matrix between each pair of K-means-generated segmentation candidates for every image at each iteration. 

Fig. \ref{fig:proposal_similarity} illustrates the evolution of representative segmentation candidate similarity across six iterations. The boxplot on the left displays the range of similarity scores across all images, while the plot on the right provides a visual representation of areas where the four correction candidates disagree. In this visualization, red areas correspond to regions identified by only one candidate as part of the target, light blue by two candidates, and purple by three candidates. 

In the first iteration, where no user feedback is involved, the segmentation candidates are relatively similar, resulting in a median similarity of approximately 0.952. This is because the model generates predictions based on the same input features, leading to only minor variations among the representative segmentation candidates. The visualization (Fig. \ref{fig:proposal_similarity}) confirms this, showing minimal divergence between candidates, with areas of disagreement being small and concentrated mainly at the optic cup boundaries. After the first human interaction (Iteration 2), the similarity between candidates decreases significantly as the user's feedback introduces new corrections based on their preference. This feedback causes the representative segmentation candidates to diverge, contributing to different plausible segmentations.
The median similarity score drops to 0.939, as shown by the noticeable increase in areas of disagreement (Fig. \ref{fig:proposal_similarity}), particularly at the optic cup boundary. The diversity in candidates at this stage reflects the model’s flexibility in generating a range of segmentations in response to user interactions. 

\begin{table*}[hbt]
\centering
\caption{\textbf{SPA Achieves Superior Dice Score Improvements Across Iterations.} Quantitative comparison of Dice Score improvements between consecutive iterations for different interactive models. The ``Overall Diff'' column shows the total Dice Score improvement from Iteration 1 to Iteration 6. SPA consistently achieves the highest performance gains, demonstrating its effectiveness in incorporating user interaction for segmentation refinement.}
\resizebox{0.65\linewidth}{!}{%
\begin{tabular}{c|cccccc}
\hline \hline
Model         & Diff 1 & Diff 2 & Diff 3 & Diff 4 & Diff 5 & Overall Diff \\ \hline
SAM \cite{kirillov_segment_2023}     & -0.03  & 0.04   & 0.14   & 0.12   & 0.13   & 0.41         \\
MedSAM \cite{ma_segment_2024}   & 0.43   & \textbf{0.45}   & 0.04   & 0.06   & 0.04   & 1.02         \\
MSA \cite{wu_medical_2023}      & 0.06   & -0.01  & 0.01   & 0.05   & -0.03  & 0.09         \\
SAM-U V1 \cite{deng_sam-u_2023} & -0.09  & -0.04  & -0.04  & 0.05   & 0.05   & -0.08        \\
SAM-U V2 \cite{deng_sam-u_2023} & -0.11  & 0.00   & 0.06   & -0.05  & 0.01   & -0.09        \\ \hline
\rowcolor{cyan!40!white!20}
SPA  & \textbf{1.05}   & 0.34   & \textbf{0.36}   & \textbf{0.16}   & \textbf{0.15}   & \textbf{2.07}       \\ \hline \hline
\end{tabular}%
}\label{tab:pred_change}
\end{table*}

\begin{figure*}[hbt!]
    \centering
    \includegraphics[scale=0.33]{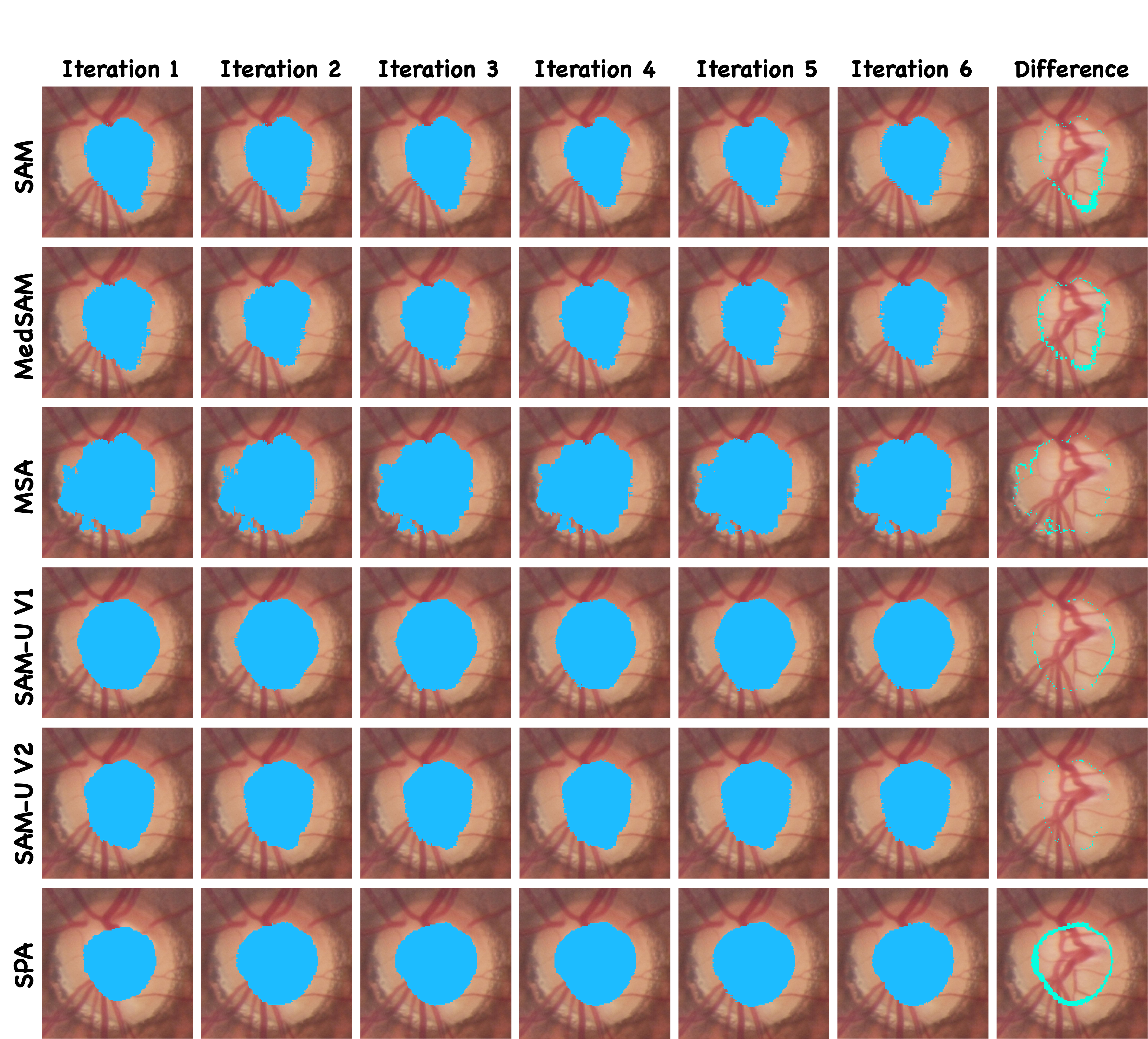
    } 
    \caption{\textbf{SPA Outperforms Other Interactive Models in Prediction Refinement.} Visual comparison of predictions from SAM, MedSAM, MSA, SAM-U (V1, V2), and SPA models across six interaction iterations. The last column shows the difference between the final and initial predictions. SPA exhibits the most significant changes between iterations, indicating its greater sensitivity to user interaction and improved refinement of segmentation predictions.}
    \label{fig:vis_pred_change}
\end{figure*}

As the human interaction process continues, our model refines its predictions based on user feedback. The correction candidates gradually converge as the model learns from the user’s interaction and moves toward a more specific, preference-aligned segmentation. This steady convergence is evident in iterations 3 through 6 in Fig. \ref{fig:proposal_similarity}, where the similarity scores gradually increase. By the final iteration, the candidates exhibit a high median similarity of 0.956, surpassing that of the initial iteration. The visualization on the right (Fig. \ref{fig:proposal_similarity}) demonstrates this convergence as the areas of disagreement shrink significantly, particularly at the boundary areas. This indicates that the model has effectively adapted to the user’s interaction, reducing uncertainty in its predictions and converging toward a consistent segmentation aligned with the human preference.

This analysis provides valuable insights into how human interactions influence the evolution of correction candidates. Initially, the candidates are similar because they are generated by the model alone, without any external corrections from user preferences. However, once the user's feedback is introduced, the candidates diverge to reflect different potential segmentations, capturing the information introduced through the user's interaction. Over time, the correction candidates begin to converge as the model refines its predictions, steadily narrowing down the range of plausible segmentations and aligning them more closely with user preferences. This steady convergence reflects the success of our multi-choice correction candidate approach, which effectively incorporates human feedback to refine the model’s predictions toward more  preference-specific segmentations.


\section{Prediction Change After Interactions}
\label{app:Prediction Change After Interactions}

In this section, we highlight the effectiveness of our SPA model's multi-choice correction candidate interaction strategy, demonstrating how it outperforms other interactive models in terms of prediction refinement. The quantitative results and visual examples illustrate how the model's sensitivity to clinician interactions leads to more substantial improvements over iterations compared to other interactive models like SAM \cite{kirillov_segment_2023}, MedSAM \cite{ma_segment_2024}, MSA \cite{wu_medical_2023}, and SAM-U variants \cite{deng_sam-u_2023}.

Table \ref{tab:pred_change} quantifies the changes in Dice Score (\%) between consecutive iterations for each model for the REFUGE2 optic cup segmentation task. ``Diff 1'' refers to the improvement from Iteration 1 to Iteration 2, ``Diff 2'' from Iteration 2 to Iteration 3, and so on. The final column shows the overall Dice Score difference between the first and last iterations for each model, serving as a cumulative measure of how much each model’s performance improved throughout the interactive process. SPA consistently achieves higher Dice Score improvements compared to all other models across almost every iteration. For example, the first interaction yields a substantial Dice Score increase of 1.05\% for SPA, whereas SAM, SAM-U, and MSA models exhibit marginal or even negative changes in performance. The overall difference for SPA is 2.07\%, significantly higher than the next-best-performing model (MedSAM with 1.02\%).

Fig. \ref{fig:vis_pred_change} shows an visual example case comparing the REFUGE2 optic cup segmentation predictions of various interactive models. Each row corresponds to a different model, while the columns display the prediction results at each interaction iteration (from Iteration 1 to Iteration 6). The final column represents the difference between the last and first predictions, highlighting how much each model's prediction has evolved due to clinician interaction. As shown in the last column in \ref{fig:vis_pred_change}, SPA exhibits the most substantial change between the first and last predictions, indicating its high sensitivity to clinician interactions. This responsiveness suggests that the SPA model is more adept at refining its predictions based on the provided feedback, leading to better alignment with the user preference. In contrast, models such as SAM, MedSAM, and MSA show more limited changes, indicating less responsiveness to the interactive corrections provided by the clinician.

The combined quantitative and qualitative evidence underscores the superior effectiveness of SPA’s multi-choice correction candidate approach. SPA is more responsive to clinician interactions, leading to larger adjustments in its predictions and greater alignment with the user preference. This higher sensitivity to clinician feedback, compared to other interactive models, results in more meaningful improvements in segmentation performance over time. SPA's capacity to integrate multiple correction candidates allows it to dynamically adjust its predictions in response to clinician interaction, enabling more effective refinement of segmentations. This adaptability makes it particularly suited for clinical environments where human interactions are critical for achieving preference-specific medical image segmentations.


\begin{figure*}[hbt!]
    \centering
    \includegraphics[scale=0.23]{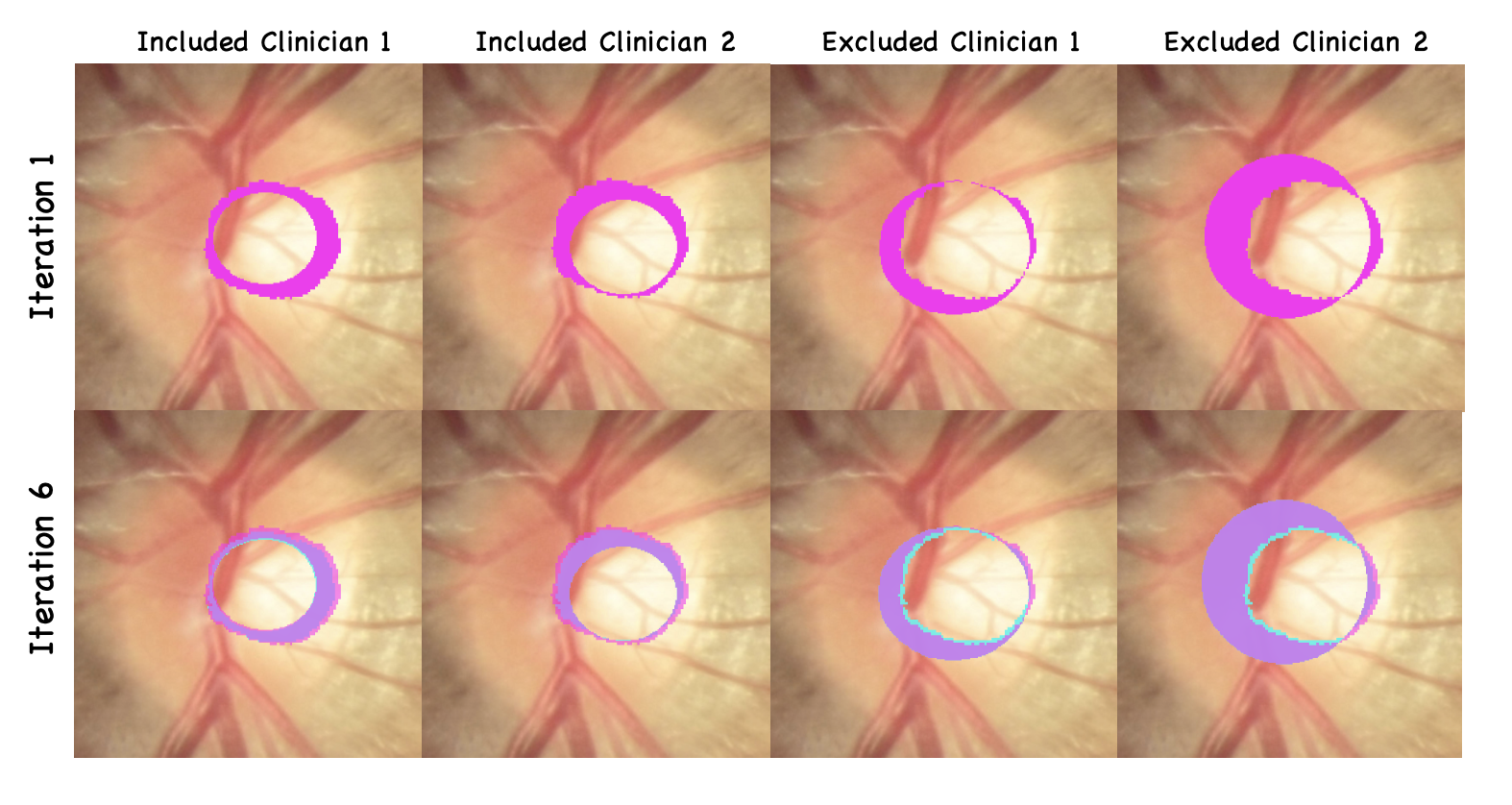
    } 
    \caption{\textbf{SPA Demonstrates Visual Prediction Alignment with Clinician Annotations.} The figure illustrates the differences between SPA’s optic cup segmentation predictions and individual clinicians' annotations in the REFUGE2 dataset across multiple interaction iterations. Dark purple represents the initial differences between the predictions and clinician annotations. Light purple indicates the overlap of differences between the first and sixth interactions, while blue shows new differences that emerge in the sixth iteration. The results demonstrate that the model increasingly aligns with included clinicians and diverges from excluded clinicians over time.}
    \label{fig:vis_pred_ali}
\end{figure*}


\section{Human User Study Details}
In order to evaluate the efficiency of our SPA model compared to the previous interactive model, MedSAM, we conducted a detailed human evaluation study simulating real-world medical image segmentation workflows. Five medical professionals, each with over five years of graduate-level expertise, participated in the study. Their task was to interact with the models to refine predictions until they met their clinical standards.

For the MedSAM model, participants were provided with two types of prompts: \textit{Click} and \textit{BBox}. These prompts could be used to include or exclude specific pixels in the target segmentation, offering flexibility for achieving desired results. Participants were allowed to select the prompt type that best suited their needs for each scenario, simulating the decision-making process in clinical practice. In contrast, the SPA model introduced a multi-choice interface designed to streamline interactions. Instead of requiring manual pixel inclusion or exclusion, SPA presented participants with several correction candidates during each iteration. This design allowed users to select the option that most closely aligned with their expectations, reducing the cognitive and manual effort involved in refining predictions.

Participants interacted with each model iteratively, making adjustments until the predictions matched their desired criteria. Throughout the study, we recorded two key metrics for each model: (1) the total time required to achieve satisfactory results and (2) the number of interaction iterations needed to reach the final output. By averaging these metrics across multiple cases, we were able to quantify and compare the efficiency of SPA and MedSAM. This evaluation demonstrated the potential of SPA’s uncertainty-aware, multi-choice framework to improve the user experience in medical image segmentation. The results suggest that SPA can significantly reduce the time and effort required to achieve accurate segmentation, while also offering greater adaptability to the needs of medical professionals.

\section{Visualization for Prediction Alignment with Clinicians} 
\label{app: Visualisation for Prediction Alignment with Clinicians}
Fig. \ref{fig:vis_pred_ali} illustrates a visual comparison of the differences between SPA's segmentation predictions and individual clinicians' annotations over six iterations. The focus is on how the model's predictions evolve with human interaction, comparing included clinicians with excluded clinicians. In Iteration 1, the dark purple areas represent the initial differences between the model's predictions and the clinicians' annotations. By Iteration 6, the light purple regions show the overlap between the differences observed in Iteration 1 and the updated differences in Iteration 6, indicating which discrepancies remain consistent across iterations. The blue areas highlight new differences introduced in Iteration 6 that are not present in the first iteration. For included clinicians, the light purple areas shrink, signifying that the model’s predictions are becoming more aligned with their annotations. In contrast, for excluded clinicians, the blue regions grow, showing that the model’s predictions are diverging from their annotations. This visualization effectively demonstrates how the model refines its predictions over interactions, aligning more closely with the annotations of included clinicians while progressively moving away from those of excluded clinicians.


%% file: main.bbl
\begin{thebibliography}{36}
\providecommand{\natexlab}[1]{#1}
\providecommand{\url}[1]{\texttt{#1}}
\expandafter\ifx\csname urlstyle\endcsname\relax
  \providecommand{\doi}[1]{doi: #1}\else
  \providecommand{\doi}{doi: \begingroup \urlstyle{rm}\Url}\fi

\bibitem[Armato et~al.(2011)Armato, McLennan, Bidaut, McNitt-Gray, Meyer, Reeves, Zhao, Aberle, Henschke, Hoffman, Kazerooni, MacMahon, van Beek, Yankelevitz, Biancardi, Bland, Brown, Engelmann, Laderach, Max, Pais, Qing, Roberts, Smith, Starkey, Batra, Caligiuri, Farooqi, Gladish, Jude, Munden, Petkovska, Quint, Schwartz, Sundaram, Dodd, Fenimore, Gur, Petrick, Freymann, Kirby, Hughes, Vande~Casteele, Gupte, Sallam, Heath, Kuhn, Dharaiya, Burns, Fryd, Salganicoff, Anand, Shreter, Vastagh, Croft, and Clarke]{armato_lung_2011}
Samuel~G. Armato, Geoffrey McLennan, Luc Bidaut, Michael~F. McNitt-Gray, Charles~R. Meyer, Anthony~P. Reeves, Binsheng Zhao, Denise~R. Aberle, Claudia~I. Henschke, Eric~A. Hoffman, Ella~A. Kazerooni, Heber MacMahon, Edwin J.~R. van Beek, David Yankelevitz, Alberto~M. Biancardi, Peyton~H. Bland, Matthew~S. Brown, Roger~M. Engelmann, Gary~E. Laderach, Daniel Max, Richard~C. Pais, David P.-Y. Qing, Rachael~Y. Roberts, Amanda~R. Smith, Adam Starkey, Poonam Batra, Philip Caligiuri, Ali Farooqi, Gregory~W. Gladish, C.~Matilda Jude, Reginald~F. Munden, Iva Petkovska, Leslie~E. Quint, Lawrence~H. Schwartz, Baskaran Sundaram, Lori~E. Dodd, Charles Fenimore, David Gur, Nicholas Petrick, John Freymann, Justin Kirby, Brian Hughes, Alessi Vande~Casteele, Sangeeta Gupte, Maha Sallam, Michael~D. Heath, Michael~H. Kuhn, Ekta Dharaiya, Richard Burns, David~S. Fryd, Marcos Salganicoff, Vikram Anand, Uri Shreter, Stephen Vastagh, Barbara~Y. Croft, and Laurence~P. Clarke.
\newblock The {Lung} {Image} {Database} {Consortium} ({LIDC}) and {Image} {Database} {Resource} {Initiative} ({IDRI}): {A} {Completed} {Reference} {Database} of {Lung} {Nodules} on {CT} {Scans}.
\newblock \emph{Medical Physics}, 38\penalty0 (2):\penalty0 915--931, 2011.

\bibitem[Ballou et~al.(1997)Ballou, Fisher, Hakala, and Farkas]{ballou_tumor_1997}
Byron Ballou, Gregory~W. Fisher, Thomas~R. Hakala, and Daniel~L. Farkas.
\newblock Tumor {Detection} and {Visualization} {Using} {Cyanine} {Fluorochrome}-{Labeled} {Antibodies}.
\newblock \emph{Biotechnology Progress}, 13\penalty0 (5):\penalty0 649--658, 1997.
\newblock \_eprint: https://onlinelibrary.wiley.com/doi/pdf/10.1021/bp970088t.

\bibitem[Baumgartner et~al.(2019)Baumgartner, Tezcan, Chaitanya, Hötker, Muehlematter, Schawkat, Becker, Donati, and Konukoglu]{baumgartner_phiseg_2019}
Christian~F. Baumgartner, Kerem~C. Tezcan, Krishna Chaitanya, Andreas~M. Hötker, Urs~J. Muehlematter, Khoschy Schawkat, Anton~S. Becker, Olivio Donati, and Ender Konukoglu.
\newblock {PHiSeg}: {Capturing} {Uncertainty} in {Medical} {Image} {Segmentation}.
\newblock In \emph{Medical {Image} {Computing} and {Computer} {Assisted} {Intervention} – {MICCAI} 2019}, pages 119--127, Cham, 2019. Springer International Publishing.

\bibitem[Cao et~al.(2021)Cao, Wang, Chen, Jiang, Zhang, Tian, and Wang]{cao_swin-unet_2021}
Hu Cao, Yueyue Wang, Joy Chen, Dongsheng Jiang, Xiaopeng Zhang, Qi Tian, and Manning Wang.
\newblock Swin-{Unet}: {Unet}-like {Pure} {Transformer} for {Medical} {Image} {Segmentation}, 2021.
\newblock arXiv:2105.05537 [cs, eess].

\bibitem[Chen et~al.(2021)Chen, Lu, Yu, Luo, Adeli, Wang, Lu, Yuille, and Zhou]{chen_transunet_2021}
Jieneng Chen, Yongyi Lu, Qihang Yu, Xiangde Luo, Ehsan Adeli, Yan Wang, Le Lu, Alan~L. Yuille, and Yuyin Zhou.
\newblock {TransUNet}: {Transformers} {Make} {Strong} {Encoders} for {Medical} {Image} {Segmentation}, 2021.
\newblock arXiv:2102.04306 [cs].

\bibitem[Chen et~al.(2022)Chen, Zhao, Zhang, Duan, Qi, and Zhao]{chen_focalclick_2022}
Xi Chen, Zhiyan Zhao, Yilei Zhang, Manni Duan, Donglian Qi, and Hengshuang Zhao.
\newblock {FocalClick}: {Towards} {Practical} {Interactive} {Image} {Segmentation}, 2022.
\newblock arXiv:2204.02574 [cs].

\bibitem[Clark et~al.(2013)Clark, Vendt, Smith, Freymann, Kirby, Koppel, Moore, Phillips, Maffitt, Pringle, Tarbox, and Prior]{clark_cancer_2013}
Kenneth Clark, Bruce Vendt, Kirk Smith, John Freymann, Justin Kirby, Paul Koppel, Stephen Moore, Stanley Phillips, David Maffitt, Michael Pringle, Lawrence Tarbox, and Fred Prior.
\newblock The {Cancer} {Imaging} {Archive} ({TCIA}): {Maintaining} and {Operating} a {Public} {Information} {Repository}.
\newblock \emph{Journal of Digital Imaging}, 26\penalty0 (6):\penalty0 1045--1057, 2013.

\bibitem[Dehnad et~al.(2014)Dehnad, Nasser, and Hosseini]{dehnad2014comparison}
Afsaneh Dehnad, Hayedeh Nasser, and Agha~Fatemeh Hosseini.
\newblock A comparison between three-and four-option multiple choice questions.
\newblock \emph{Procedia-Social and Behavioral Sciences}, 98:\penalty0 398--403, 2014.

\bibitem[Deng et~al.(2023)Deng, Zou, Ren, Wang, Yuan, Ying, and Fu]{deng_sam-u_2023}
Guoyao Deng, Ke Zou, Kai Ren, Meng Wang, Xuedong Yuan, Sancong Ying, and Huazhu Fu.
\newblock {SAM}-{U}: {Multi}-box prompts triggered uncertainty estimation for reliable {SAM} in medical image, 2023.
\newblock arXiv:2307.04973 [cs].

\bibitem[Fang et~al.(2022)Fang, Li, Wu, Fu, Sun, Son, Yu, Zhang, Yuan, Bian, Lei, Zhao, Xu, Li, Fumero, Sigut, Almubarak, Bazi, Guo, Zhou, Baid, Innani, Guo, Yang, Orlando, Bogunović, Zhang, and Xu]{fang_refuge2_2022}
Huihui Fang, Fei Li, Junde Wu, Huazhu Fu, Xu Sun, Jaemin Son, Shuang Yu, Menglu Zhang, Chenglang Yuan, Cheng Bian, Baiying Lei, Benjian Zhao, Xinxing Xu, Shaohua Li, Francisco Fumero, José Sigut, Haidar Almubarak, Yakoub Bazi, Yuanhao Guo, Yating Zhou, Ujjwal Baid, Shubham Innani, Tianjiao Guo, Jie Yang, José~Ignacio Orlando, Hrvoje Bogunović, Xiulan Zhang, and Yanwu Xu.
\newblock {REFUGE2} {Challenge}: {A} {Treasure} {Trove} for {Multi}-{Dimension} {Analysis} and {Evaluation} in {Glaucoma} {Screening}, 2022.
\newblock arXiv:2202.08994 [cs, eess].

\bibitem[Guan et~al.(2018)Guan, Gulshan, Dai, and Hinton]{guan_who_2018}
Melody~Y. Guan, Varun Gulshan, Andrew~M. Dai, and Geoffrey~E. Hinton.
\newblock Who {Said} {What}: {Modeling} {Individual} {Labelers} {Improves} {Classification}, 2018.
\newblock arXiv:1703.08774 [cs].

\bibitem[He et~al.(2021)He, Chen, Xie, Li, Dollár, and Girshick]{he_masked_2021}
Kaiming He, Xinlei Chen, Saining Xie, Yanghao Li, Piotr Dollár, and Ross Girshick.
\newblock Masked {Autoencoders} {Are} {Scalable} {Vision} {Learners}, 2021.
\newblock arXiv:2111.06377 [cs].

\bibitem[Hesamian et~al.(2019)Hesamian, Jia, He, and Kennedy]{hesamian_deep_2019}
Mohammad~Hesam Hesamian, Wenjing Jia, Xiangjian He, and Paul Kennedy.
\newblock Deep {Learning} {Techniques} for {Medical} {Image} {Segmentation}: {Achievements} and {Challenges}.
\newblock \emph{Journal of Digital Imaging}, 32\penalty0 (4):\penalty0 582--596, 2019.

\bibitem[Huang et~al.(2024)Huang, Ruan, Xing, and Feng]{huang_review_2024}
Ling Huang, Su Ruan, Yucheng Xing, and Mengling Feng.
\newblock A review of uncertainty quantification in medical image analysis: {Probabilistic} and non-probabilistic methods.
\newblock \emph{Medical Image Analysis}, 97:\penalty0 103223, 2024.

\bibitem[Jensen et~al.(2019)Jensen, Jørgensen, Jalaboi, Hansen, and Olsen]{jensen_improving_2019}
Martin~Holm Jensen, Dan~Richter Jørgensen, Raluca Jalaboi, Mads~Eiler Hansen, and Martin~Aastrup Olsen.
\newblock Improving {Uncertainty} {Estimation} in {Convolutional} {Neural} {Networks} {Using} {Inter}-rater {Agreement}.
\newblock In \emph{Medical {Image} {Computing} and {Computer} {Assisted} {Intervention} – {MICCAI} 2019}, pages 540--548, Cham, 2019. Springer International Publishing.

\bibitem[Ji et~al.(2021)Ji, Yu, Wu, Ma, Bian, Bi, Li, Liu, Cheng, and Zheng]{ji_learning_2021}
Wei Ji, Shuang Yu, Junde Wu, Kai Ma, Cheng Bian, Qi Bi, Jingjing Li, Hanruo Liu, Li Cheng, and Yefeng Zheng.
\newblock Learning {Calibrated} {Medical} {Image} {Segmentation} via {Multi}-rater {Agreement} {Modeling}.
\newblock In \emph{2021 {IEEE}/{CVF} {Conference} on {Computer} {Vision} and {Pattern} {Recognition} ({CVPR})}, pages 12336--12346, 2021.
\newblock ISSN: 2575-7075.

\bibitem[Kendall and Gal(2017)]{kendall_what_2017}
Alex Kendall and Yarin Gal.
\newblock What {Uncertainties} {Do} {We} {Need} in {Bayesian} {Deep} {Learning} for {Computer} {Vision}?, 2017.
\newblock arXiv:1703.04977 [cs].

\bibitem[Kirillov et~al.(2023)Kirillov, Mintun, Ravi, Mao, Rolland, Gustafson, Xiao, Whitehead, Berg, Lo, Dollár, and Girshick]{kirillov_segment_2023}
Alexander Kirillov, Eric Mintun, Nikhila Ravi, Hanzi Mao, Chloe Rolland, Laura Gustafson, Tete Xiao, Spencer Whitehead, Alexander~C. Berg, Wan-Yen Lo, Piotr Dollár, and Ross Girshick.
\newblock Segment {Anything}, 2023.
\newblock arXiv:2304.02643 [cs].

\bibitem[Kiureghian and Ditlevsen(2009)]{kiureghian_aleatory_2009}
Armen~Der Kiureghian and Ove Ditlevsen.
\newblock Aleatory or epistemic? {Does} it matter?
\newblock \emph{Structural Safety}, 31\penalty0 (2):\penalty0 105--112, 2009.

\bibitem[Kohl et~al.(2018)Kohl, Romera-Paredes, Meyer, De~Fauw, Ledsam, Maier-Hein, Eslami, Rezende, and Ronneberger]{kohl_probabilistic_2018}
Simon A.~A. Kohl, Bernardino Romera-Paredes, Clemens Meyer, Jeffrey De~Fauw, Joseph~R. Ledsam, Klaus~H. Maier-Hein, S.~M.~Ali Eslami, Danilo~Jimenez Rezende, and Olaf Ronneberger.
\newblock A {Probabilistic} {U}-{Net} for {Segmentation} of {Ambiguous} {Images}, 2018.

\bibitem[Kohl et~al.(2019{\natexlab{a}})Kohl, Romera-Paredes, Maier-Hein, Rezende, Eslami, Kohli, Zisserman, and Ronneberger]{kohl_hierarchical_2019}
Simon A.~A. Kohl, Bernardino Romera-Paredes, Klaus~H. Maier-Hein, Danilo~Jimenez Rezende, S.~M.~Ali Eslami, Pushmeet Kohli, Andrew Zisserman, and Olaf Ronneberger.
\newblock A {Hierarchical} {Probabilistic} {U}-{Net} for {Modeling} {Multi}-{Scale} {Ambiguities}, 2019{\natexlab{a}}.
\newblock arXiv:1905.13077 [cs].

\bibitem[Kohl et~al.(2019{\natexlab{b}})Kohl, Romera-Paredes, Meyer, De~Fauw, Ledsam, Maier-Hein, Eslami, Rezende, and Ronneberger]{kohl_probabilistic_2019}
Simon A.~A. Kohl, Bernardino Romera-Paredes, Clemens Meyer, Jeffrey De~Fauw, Joseph~R. Ledsam, Klaus~H. Maier-Hein, S.~M.~Ali Eslami, Danilo~Jimenez Rezende, and Olaf Ronneberger.
\newblock A {Probabilistic} {U}-{Net} for {Segmentation} of {Ambiguous} {Images}, 2019{\natexlab{b}}.
\newblock arXiv:1806.05034 [cs, stat].

\bibitem[Li et~al.(2024)Li, Navarro, Ezhov, Bayat, Das, Kofler, Shit, Waldmannstetter, Paetzold, Hu, Wiestler, Zimmer, Amiranashvili, Prabhakar, Berger, Weidner, Alonso-Basant, Rashid, Baid, Adel, Ali, Baheti, Bai, Bhatt, Cetindag, Chen, Cheng, Dutand, Dular, Elattar, Feng, Gao, Huisman, Hu, Innani, Jiat, Karimi, Kuijf, Kwak, Le, Lia, Lin, Liu, Ma, Ma, Ma, Oksuz, Holland, Oliveira, Pal, Pei, Qiao, Saha, Selvan, Shen, Silva, Spiclin, Talbar, Wang, Wang, Wang, Wang, Xia, Xu, Yan, Yergin, Yu, Zeng, Zhang, Zhao, Zheng, Zukovec, Do, Becker, Simpson, Konukoglu, Jakab, Bakas, Joskowicz, and Menze]{li_qubiq_2024}
Hongwei~Bran Li, Fernando Navarro, Ivan Ezhov, Amirhossein Bayat, Dhritiman Das, Florian Kofler, Suprosanna Shit, Diana Waldmannstetter, Johannes~C. Paetzold, Xiaobin Hu, Benedikt Wiestler, Lucas Zimmer, Tamaz Amiranashvili, Chinmay Prabhakar, Christoph Berger, Jonas Weidner, Michelle Alonso-Basant, Arif Rashid, Ujjwal Baid, Wesam Adel, Deniz Ali, Bhakti Baheti, Yingbin Bai, Ishaan Bhatt, Sabri~Can Cetindag, Wenting Chen, Li Cheng, Prasad Dutand, Lara Dular, Mustafa~A. Elattar, Ming Feng, Shengbo Gao, Henkjan Huisman, Weifeng Hu, Shubham Innani, Wei Jiat, Davood Karimi, Hugo~J. Kuijf, Jin~Tae Kwak, Hoang~Long Le, Xiang Lia, Huiyan Lin, Tongliang Liu, Jun Ma, Kai Ma, Ting Ma, Ilkay Oksuz, Robbie Holland, Arlindo~L. Oliveira, Jimut~Bahan Pal, Xuan Pei, Maoying Qiao, Anindo Saha, Raghavendra Selvan, Linlin Shen, Joao~Lourenco Silva, Ziga Spiclin, Sanjay Talbar, Dadong Wang, Wei Wang, Xiong Wang, Yin Wang, Ruiling Xia, Kele Xu, Yanwu Yan, Mert Yergin, Shuang Yu, Lingxi Zeng, YingLin Zhang, Jiachen Zhao, Yefeng
  Zheng, Martin Zukovec, Richard Do, Anton Becker, Amber Simpson, Ender Konukoglu, Andras Jakab, Spyridon Bakas, Leo Joskowicz, and Bjoern Menze.
\newblock {QUBIQ}: {Uncertainty} {Quantification} for {Biomedical} {Image} {Segmentation} {Challenge}, 2024.
\newblock arXiv:2405.18435 [cs, eess].

\bibitem[Liu et~al.(2023)Liu, Xu, Bertasius, and Niethammer]{liu_simpleclick_2023}
Qin Liu, Zhenlin Xu, Gedas Bertasius, and Marc Niethammer.
\newblock {SimpleClick}: {Interactive} {Image} {Segmentation} with {Simple} {Vision} {Transformers}, 2023.
\newblock arXiv:2210.11006 [cs].

\bibitem[Ma et~al.(2024)Ma, He, Li, Han, You, and Wang]{ma_segment_2024}
Jun Ma, Yuting He, Feifei Li, Lin Han, Chenyu You, and Bo Wang.
\newblock Segment {Anything} in {Medical} {Images}.
\newblock \emph{Nature Communications}, 15\penalty0 (1):\penalty0 654, 2024.
\newblock arXiv:2304.12306 [cs, eess].

\bibitem[Prasanna et~al.(2012)Prasanna, Stone, Wong, Capala, Bernhard, Vikram, and Coleman]{prasanna_normal_2012}
Pataje~G.S. Prasanna, Helen~B. Stone, Rosemary~S. Wong, Jacek Capala, Eric~J. Bernhard, Bhadrasain Vikram, and C.~N. Coleman.
\newblock Normal tissue protection for improving radiotherapy: {Where} are the {Gaps}?
\newblock \emph{Translational cancer research}, 1\penalty0 (1):\penalty0 35--48, 2012.

\bibitem[Ronneberger et~al.(2015)Ronneberger, Fischer, and Brox]{ronneberger_u-net_2015}
Olaf Ronneberger, Philipp Fischer, and Thomas Brox.
\newblock U-{Net}: {Convolutional} {Networks} for {Biomedical} {Image} {Segmentation}, 2015.
\newblock arXiv:1505.04597 [cs].

\bibitem[Rupprecht et~al.(2017)Rupprecht, Laina, DiPietro, Baust, Tombari, Navab, and Hager]{rupprecht_learning_2017}
Christian Rupprecht, Iro Laina, Robert DiPietro, Maximilian Baust, Federico Tombari, Nassir Navab, and Gregory~D. Hager.
\newblock Learning in an {Uncertain} {World}: {Representing} {Ambiguity} {Through} {Multiple} {Hypotheses}, 2017.
\newblock arXiv:1612.00197 [cs].

\bibitem[Sakinis et~al.(2019)Sakinis, Milletari, Roth, Korfiatis, Kostandy, Philbrick, Akkus, Xu, Xu, and Erickson]{sakinis_interactive_2019}
Tomas Sakinis, Fausto Milletari, Holger Roth, Panagiotis Korfiatis, Petro Kostandy, Kenneth Philbrick, Zeynettin Akkus, Ziyue Xu, Daguang Xu, and Bradley~J. Erickson.
\newblock Interactive segmentation of medical images through fully convolutional neural networks, 2019.
\newblock arXiv:1903.08205 [cs].

\bibitem[Sofiiuk et~al.(2021)Sofiiuk, Petrov, and Konushin]{sofiiuk_reviving_2021}
Konstantin Sofiiuk, Ilia~A. Petrov, and Anton Konushin.
\newblock Reviving {Iterative} {Training} with {Mask} {Guidance} for {Interactive} {Segmentation}, 2021.
\newblock arXiv:2102.06583 [cs].

\bibitem[Tajbakhsh et~al.(2016)Tajbakhsh, Shin, Gurudu, Hurst, Kendall, Gotway, and Liang]{tajbakhsh_convolutional_2016}
Nima Tajbakhsh, Jae~Y. Shin, Suryakanth~R. Gurudu, R.~Todd Hurst, Christopher~B. Kendall, Michael~B. Gotway, and Jianming Liang.
\newblock Convolutional {Neural} {Networks} for {Medical} {Image} {Analysis}: {Full} {Training} or {Fine} {Tuning}?
\newblock \emph{IEEE Transactions on Medical Imaging}, 35\penalty0 (5):\penalty0 1299--1312, 2016.
\newblock arXiv:1706.00712 [cs].

\bibitem[Wang et~al.(2018)Wang, Li, Zuluaga, Pratt, Patel, Aertsen, Doel, David, Deprest, Ourselin, and Vercauteren]{wang_interactive_2018}
Guotai Wang, Wenqi Li, Maria~A. Zuluaga, Rosalind Pratt, Premal~A. Patel, Michael Aertsen, Tom Doel, Anna~L. David, Jan Deprest, Sebastien Ourselin, and Tom Vercauteren.
\newblock Interactive {Medical} {Image} {Segmentation} using {Deep} {Learning} with {Image}-specific {Fine}-tuning.
\newblock \emph{IEEE Transactions on Medical Imaging}, 37\penalty0 (7):\penalty0 1562--1573, 2018.
\newblock arXiv:1710.04043 [cs].

\bibitem[Wu et~al.(2023)Wu, Ji, Liu, Fu, Xu, Xu, and Jin]{wu_medical_2023}
Junde Wu, Wei Ji, Yuanpei Liu, Huazhu Fu, Min Xu, Yanwu Xu, and Yueming Jin.
\newblock Medical {SAM} {Adapter}: {Adapting} {Segment} {Anything} {Model} for {Medical} {Image} {Segmentation}, 2023.
\newblock arXiv:2304.12620 [cs].

\bibitem[Zhang et~al.(2020)Zhang, Tanno, Xu, Jin, Jacob, Ciccarelli, Barkhof, and Alexander]{zhang_disentangling_2020}
Le Zhang, Ryutaro Tanno, Mou-Cheng Xu, Chen Jin, Joseph Jacob, Olga Ciccarelli, Frederik Barkhof, and Daniel~C. Alexander.
\newblock Disentangling {Human} {Error} from the {Ground} {Truth} in {Segmentation} of {Medical} {Images}, 2020.
\newblock arXiv:2007.15963 [cs].

\bibitem[Zhu et~al.(2024)Zhu, Qi, and Wu]{zhu_medical_2024}
Jiayuan Zhu, Yunli Qi, and Junde Wu.
\newblock Medical {SAM} 2: {Segment} medical images as video via {Segment} {Anything} {Model} 2, 2024.
\newblock arXiv:2408.00874 [cs].

\bibitem[Zou et~al.(2023)Zou, Chen, Yuan, Shen, Wang, and Fu]{zou_review_2023}
Ke Zou, Zhihao Chen, Xuedong Yuan, Xiaojing Shen, Meng Wang, and Huazhu Fu.
\newblock A {Review} of {Uncertainty} {Estimation} and its {Application} in {Medical} {Imaging}, 2023.
\newblock arXiv:2302.08119 [cs, eess].

\end{thebibliography}
